 \documentclass[preprint2]{aastex}

\shorttitle{Meteorologic parameters analysis above Dome C - Antarctic Plateau}
\shortauthors{}

\begin{document}



\title{Meteorologic parameters analysis above \\ Dome C made with ECMWF data}



\author{Kerstin Gei{\ss}ler}
\affil{European Southern Observatory, Alonso de Cordova 3107, Santiago, Chile \\ 
Max-Planck Institut f\"ur Astronomie, Konigstuhl 17 - D69117, Heidelberg, Germany}
\author{Elena Masciadri}
\affil{INAF - Osservatorio Astrofisico di Arcetri, L.go E. Fermi 5, 50125 Florence, Italy \\ 
Max-Planck Institut f\"ur Astronomie, Konigstuhl 17 - D69117, Heidelberg, Germany}
\email{masciadri@arcetri.astro.it}


\def\LO{\mbox{${\cal L}_0 \ $}}
\def\CN2{\mbox{$C_N^2 \ $}}
\def\CNb{\mbox{$C_{N,2}^2 \ $}}
\def\CNd{\mbox{$C_{N,1}^2 \ $}}
\def\CT2{\mbox{$C_T^2 \ $}}
\def\tauAO{\mbox{$\tau_{AO} \ $}}
\def\taus{\mbox{$\tau_s \ $}}
\def\thetaAO{\mbox{$\theta_{AO} \ $}}
\def\see{\mbox{$\varepsilon\ $}}
\def\seeFA{\mbox{$\varepsilon_{FA} \ $}}
\def\seeBL{\mbox{$\varepsilon_{BL} \ $}}
\def\seed{\mbox{$\varepsilon_{d} \ $}}
\def\seesurf{\mbox{$\varepsilon_{surf.} \ $}}
\def\seem{\mbox{$\varepsilon_{m} \ $}}
\def\seeTOT{\mbox{$\varepsilon_{TOT} \ $}}
\def\sigmaI2{\mbox{$\sigma ^{2}_{I} \ $}}
\def\ventvet{\mbox{$\mathbf{\vec{V}} \ $}}
\begin{abstract}
In this paper we present the characterization of all the principal meteorological parameters (wind speed and direction, pressure, absolute and potential temperature) extended up to $25$ km from the ground and over two years (2003 and 2004) above the Antarctic site of Dome C. The data set is composed by {\it 'analyses'} provided by the General Circulation Model (GCM) of the European Center for Medium Weather Forecasts (ECMWF) and they are part of the catalog MARS. A monthly and seasonal (summer and winter time) statistical analysis of the results is presented. The Richardson number is calculated for each month of the year over $25$ km to study the stability/instability of the atmosphere. This permits us to trace a map indicating where and when the optical turbulence has the highest probability to be triggered on the whole troposphere, tropopause and stratosphere. We finally try to predict the best expected isoplanatic angle and wavefront coherence time ($\theta_{0,max}$ and a $\tau_{0,max}$) employing the Richardson number maps, the wind speed profiles and simple analytical models of $\CN2$ vertical profiles.
\end{abstract}





\section{Introduction}

The Antarctic Plateau has revealed to be particularly attractive for astronomy since 
already several years Fossat (2005), Storey et al. (2003). 
It is extremely cold and dry and this does of this site an interesting 
candidate for astronomy in the
long wavelength ranges (infrared, sub-millimeter and millimeter) thanks to the 
low sky brightness 
and high atmospheric transmission caused by a low temperature and concentration of the water vapour 
in the atmosphere (Valenziano \& Dall'Oglio 1999, Lawrence 2004, Walden et al. 2005). 
The Antarctic Plateau is placed at high altitudes (the whole continent has an average height 
of $\sim$ $2500$ m), it is characterized by a quite peculiar atmospheric circulation and a quite stable atmosphere so that the level of the optical turbulence ($\CN2$ profiles) in the free atmosphere
is, for most of the time, lower than above whatever other mid-latitude sites (Marks et al. 1996, Marks et al. 1999, Aristidi et al. 2003, Lawrence et al. 2004). Gillingham (1991), suggested for the first time, such a low level of the optical turbulence above the Antarctic Plateau. Atmospheric conditions, in general, degrade in proximity of the coasts. 
A low level of optical turbulence in the free atmosphere is, in general, associated to large isoplanatic angles ($\theta_{0}$). The coherence wavefront time ($\tau_{0}$) is claimed to be particularly large above the Antarctic Plateau due to the combination of a weak $\CN2$ and a low wind speed all along the whole troposphere. Under these conditions, an adaptive optics system can reach better levels of correction (minor residual wavefront perturbations) than those obtained by an equivalent AO system above mid-latitude sites. Wavefront correction at high Zernike orders can be more easily reached over a large field of view, the wavefront-corrector can run at reasonably low frequencies and observations with long exposure time can be done in closed loop. This could reveal particularly advantageous for some scientific programs such as searches for extra-solar planets. Of course, also the interferometry would benefit from a weak $\CN2$ and $\tau_{0}$. \newline

In the last decade several site testing campaigns took place, first above South Pole (Marks et al., 1996, Loewenstein et al. 1998, Marks et al. 1999, Travouillon et al. 2003a, Travouillon 2003b) and, more recently, above Dome C (Aristidi et al. 2003, Aristidi et al. 2005a, Lawrence et al. 2004). Dome C seems to have some advantages with respect to the South Pole: {\bf (a)} The sky emission and atmospheric transparency is some order of magnitude better than above South Pole (Lawrence 2004) at some wavelengths. The sensitivity (depending on the decreasing of sky emission and increasing of transparency) above Dome C is around 2 times better than above South Pole in near to mid-infrared regions and around 10 times better in mid to far-infrared regions. {\bf (b)} the surface turbulent layer, principally originated by the katabatic winds, is much more thinner above Dome C (tens of meters - Aristidi et al. 2005a, Lawrence et al. 2004) than above South Pole (hundreds of meters - Marks et al. 1999). The thickness and strength of the surface turbulent layer is indeed tightly correlated to the katabatic winds, a particular wind developed near the ground characterizing the boundary layer circulation above the whole Antarctic continent. Katabatic winds are produced by the radiative cooling of the iced surface that, by conduction, cools the air in its proximity. The cooled air, in proximity of the surface, becomes heavier than the air in the up layers and, for a simple gravity effect, it moves down following the ground slope with a speed increasing with the slope. Dome C is located on the top of an Altiplano in the interior region of Antarctica and, for this reason, the katabatic winds are much weaker above Dome C than above other sites in this continent such as South Pole placed on a more accentuated sloping region. \newline

At present not much is known about the typical values of meteorological parameters above Dome C during the winter (April-September) time i.e. the most interesting period for astronomers.

\noindent
The goals of our study are the following.\newline 
{\bf (i)} We intend to provide a complete analysis of the
vertical distribution of the main meteorological parameters (wind speed and direction, absolute temperature, pressure) in different months of the year
using European Center For Medium Weather Forecasts (ECMWF) data. A particular attention is addressed to the wind speed, key element for the estimate of the wavefront coherence time $\tau_{0}$.
The ECMWF data-set is produced by the ECMWF General Circulation Model (GCM) and is therefore
reliable at synoptic scale i.e. at large spatial scale. This means that 
our analysis can be extended to the whole troposphere and even stratosphere 
up to $20$-$25$ km. 
The accuracy of such a kind of data is not particularly high in the first meters above the ground due to 
the fact that the orographic effects produced by the friction of the atmospheric flow above the ground are not necessarily well reconstructed by 
the GCMs\footnote{It was recently shown (Masciadri 2003) that meso-scale models can estimate the near ground wind above astronomical sites better than the GCMs.}. 
We remind to the reader that a detailed analysis of the wind speed near the ground above Dome C extended over a time scale of $20$ years was recently presented by Aristidi et al. (2005a). In that paper, measurements of wind speed taken with an automatic weather station (AWS) ($http://uwamrc.ssec.wisc.edu$) are used to characterize the typical climatological trend of this parameter. In the same paper it is underlined that estimates of the temperature near the ground are provided by Schwerdtfeger (1984). The interested reader can find information on the value of this meteorologic parameter above Dome C and near the surface in these references. Our analysis can therefore complete the picture providing typical values (seasonal trend and median values) of the meteorological parameters in the high part of the surface layer, the boundary layer and the free atmosphere. Thanks to the large and homogeneous temporal coverage of ECMWF data we 
will be able to put in evidence typical features of the meteorological parameters in the summer and winter time and the variability of the meteorological parameters in different years. The winter time is particularly attractive for astronomical applications due to the persistence of the {\it 'night time'} for several months. This period is also the one in which it is more difficult to carry out measurements of meteorological parameters due to logistic problems. For this reason ECMWF data offer a useful alternative to measurements for monitoring the atmosphere above Dome C over long time scales in the future. \newline\newline 
{\bf (ii)} We intend to study the conditions of stability/instability of the atmosphere that can be measured by the Richardson number that depends on both the gradient of the potential temperature and the wind speed: $R_{i}$$=$$R_{i}$(${\partial \theta }/{\partial h}$,${\partial V}/{\partial h}$). The trigger of optical turbulence in the atmosphere depends on both the gradient of the potential temperature (${\partial \theta }/{\partial h}$) and the wind speed (${\partial V}/{\partial h}$) i.e. from the $R_{i}$. This parameter can therefore provide useful information on the probability to find turbulence at different altitudes in the troposphere and stratosphere in different period of the year. Why this is interesting ? At present we have indications that, above Dome C, the optical turbulence is concentrated in a thin surface layer. Above this layer the $r_{0}$ is exceptionally large indicating an extremely low level of turbulence. The astronomic community collected so far several elements certifying the excellent quality of the Dome C site and different solutions might be envisaged to overcome the strong surface layer such as rising up a telescope above $30$ m or compensating for the surface layer with AO techniques. The challenging question is now to establish more precisely how much the Dome C is better than a mid-latitude site. In other words, which are the {\it typical} $\varepsilon$, $\tau_{0}$ and $\theta_{0}$ that we can expect from this site ?  We mean here as {\it typical}, values that repeat with a statistical relevance such as a mean or a median value. For example, the gain in terms of impact on instrumentation performances and astrophysical feedback can strongly change depending on how weak the $\CN2$ is above the first $30$ m. In spite of the fact that $\CN2$$=$$10^{-18}$, $\CN2$$=$$10^{-19}$ or $\CN2$$=$$0$ are all small quantities, they can have a different impact on the final value of $\varepsilon$, $\tau_{0}$ and $\theta_{0}$. Only a precise estimate of this parameter will provide to the astronomic community useful elements to better plan future facilities (telescopes or interferometers) above the Antarctic Plateau and to correctly evaluate the real advantage in terms of turbulence obtained choosing the Antarctic Plateau as astronomical site. With the support of the Richardson number, the wind speed profile and a simple analytical $\CN2$ model we will try to predict a $\tau_{0,max}$ and a $\theta_{0,max}$ without the contribution of the first $30$ m of atmosphere. \newline\newline
{\bf (iii)} Data provided by ECMWF can be used as inputs for atmospheric meso-scale models usually employed to simulate the optical turbulence ($\CN2$) and the integrated astroclimatic parameters (Masciadri et al. 2004, Masciadri \& Egner 2004, Masciadri \& Egner 2005). Measurements of wind speed done during the summer time have been recently published (Fig. 1 - Aristidi et al. 2005a). We intend to estimate the quality and reliability of the ECMWF data comparing these values with measurements from Aristidi et al. so to have an indication of the quality of the initialization data for meso-scale models. We planned applications of a meso-scale model 
(Meso-Nh) to the Dome C in the near-future. As a further output this model will be able to reconstruct, in a more accurate way than the ECMWF data-set, the meteorologic parameters near the ground.\newline

The paper is organized in the following way. In Section \ref{meteo} we present the median values of the main meteorological parameters and their seasonal trend. We also present a study of the Richardson number tracing a complete map of the instability/stability regions in the whole $25$ km on a monthly statistical base. In Section \ref{rel} we study the reliability of our estimate comparing ECMWF analysis with measurements. In Section \ref{disc} we try to retrieve the typical value of  $\tau_{0,max}$ and $\theta_{0,max}$ above Dome C. Finally, in Section \ref{conc} we present our conclusions.

\section{Meteorological Parameters Analysis}
\label{meteo}
\begin{figure*}
\centering
\includegraphics[width=12cm,angle=90]{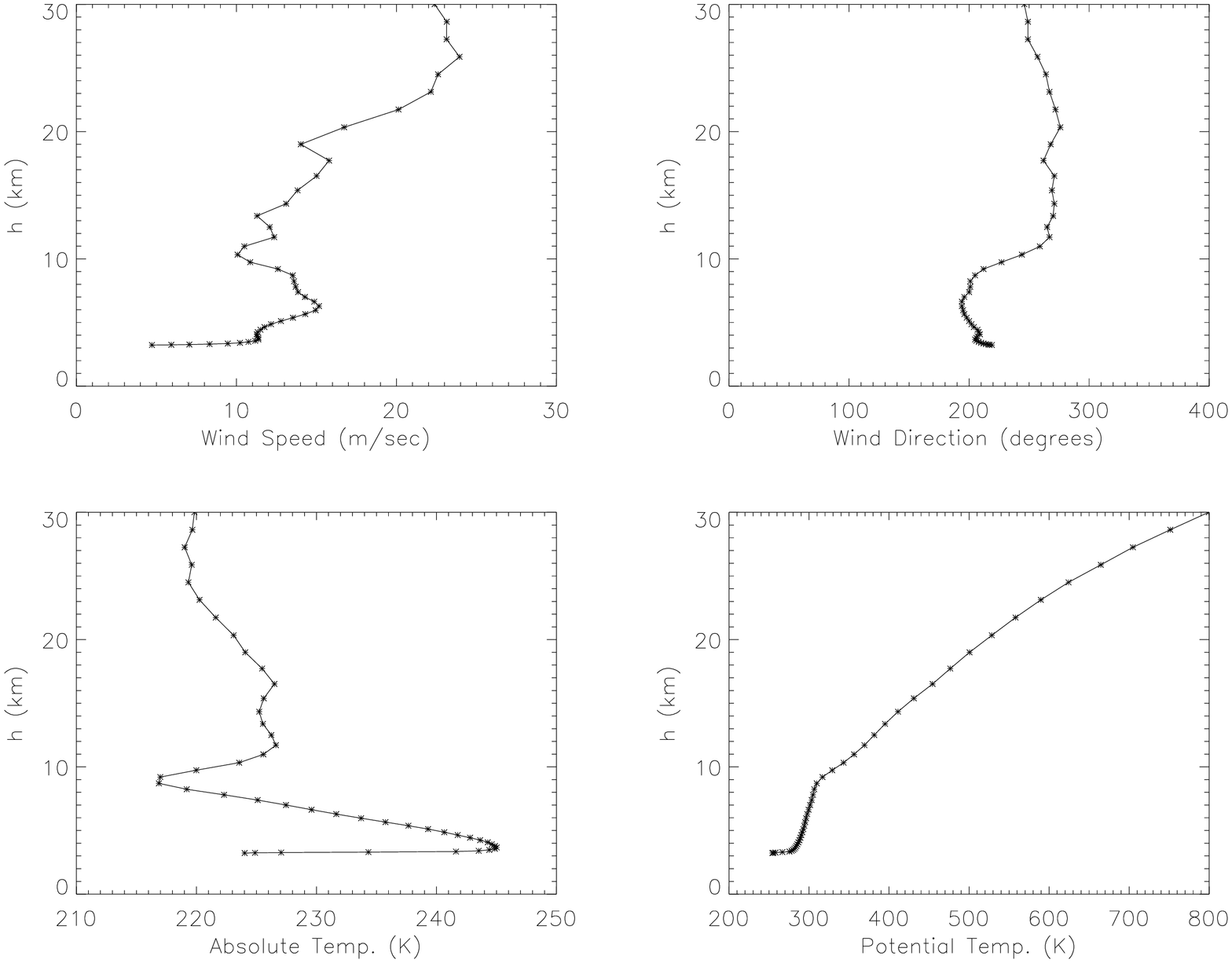}
\caption{Vertical profiles of wind speed, wind direction, absolute and potential temperature in the format of the MARS catalog (ECMWF archive). The asterisk indicate the spatial sampling over which the values of the meteorologic parameters are delivered.
\label{mars}}
\end{figure*}

The characterization of the meteorological parameters is done in this paper with {\it 'analyses'} extracted by the catalog MARS (Meteorological Archival and Retrieval System) of the ECMWF. An {\it 'analysis'} provided by the ECMWF general circulation (GCM) model is the output of a calculation based on a set of spatio-temporal interpolations of measurements provided by meteorological stations distributed on the surface of the whole world and by satellite as well as instruments carried aboard aircrafts. These measurements are continuously up-dated and the model is fed by new measurements at regular intervals of few hours. The outputs are formed by a set of fields (scalar and/or vectors) of classical meteorological parameters sampled on the whole world with a horizontal resolution of $0.5$$^{\circ}$ correspondent to roughly $50$ km. This horizontal resolution is quite better than that of the NCEP/NCAR Reanalyses having an horizontal resolution of $2.5$$^{\circ}$ so we can expect more accurate estimate of the meteorological parameters in the atmosphere. The vertical profiles are sampled over $60$ levels extended up to $60$ km. The vertical resolution is higher near the ground ($\sim$ $15$ m above Dome C) and weaker in the high part of the atmosphere. In order to give an idea of the vertical sampling, Fig. \ref{mars} shows the output of one data-set (wind speed and direction, absolute and potential temperature) of the MARS catalog (extended in the first $30$ km) with the correspondent levels at which estimates are provided. We extracted from the ECMWF archive a vertical profile of all the most important meteorological parameters (wind speed and direction, pressure, absolute and potential temperature) in the coordinates (75$^{\circ}$ S, 123$^{\circ}$ E) at $00$:$00$ U.T. for each day of the 2003 and 2004 years. We verified that the vertical profiles of the meteorologic parameters extracted from the nearest 4 grid points around the Dome C (75$^{\circ}$$06^{\prime}$$25''$ S, 123$^{\circ}$$20^{\prime}$$44''$ E) show negligible differences. This is probably due to the fact that the orography of the Antarctic continent is quite smoothed and flat in proximity of Dome C. Above this site we can appreciate on an orographic map a difference in altitude of the order of a few meters over a surface of $60$ kilometers (Masciadri 2000), roughly the distance between 2 contiguous grid points of the GCM. The orographic effects on the atmospheric flow are visibly weak at such a large spatial scale on the whole $25$ km. We can therefore consider that these profiles of meteorologic parameters at macroscopic scale well represent the atmospheric characteristics above Dome C starting from the first ten of meters as previously explained.   

\subsection{Wind speed}

\begin{figure*}
\centering
\includegraphics[width=16cm]{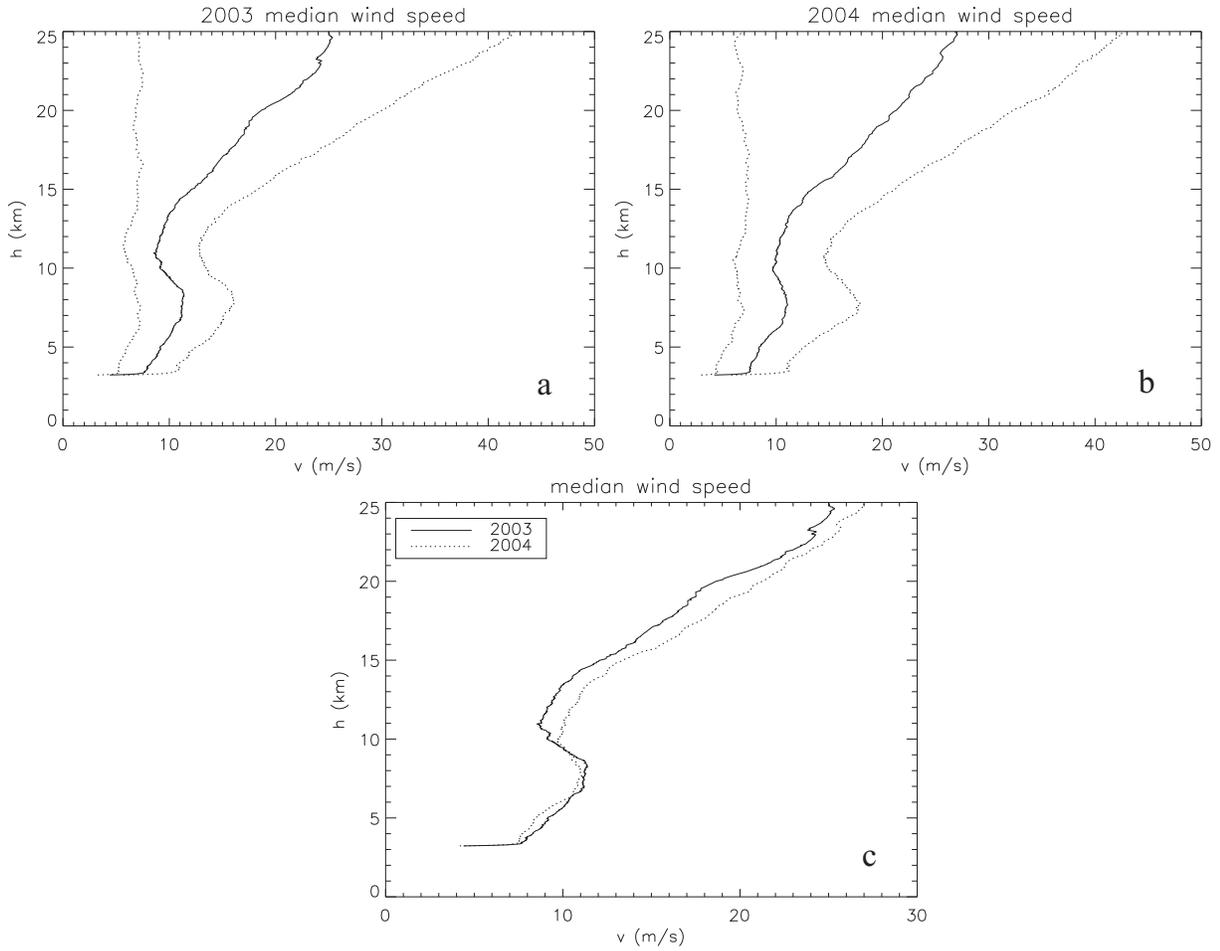}
\caption{Yearly median wind speed vertical profile. (a)-(b) Median wind speed vertical profile (full line) and first and third quartiles (dotted lines) during the 2003 and 2004 years. (c) Median wind speed vertical profiles during the 2003 and 2004 years.
\label{year_wind}}
\end{figure*}

The wind speed is one among the most critical parameters defining the quality of an astronomical site. It plays a fundamental role in triggering optical turbulence ($\CN2$) and it is a fundamental parameter in the definition of the wavefront coherence time $\tau_{0}$:

\begin{equation}
\tau _{0}=0.049\cdot \lambda ^{6/5}\left[ \int V\left( h\right) ^{5/3}\cdot
C_{N}^{2}\left( h\right) dh\right] ^{-3/5}
\label{eq1}
\end{equation}
\noindent
where $\lambda$ is the wavelength, V the wind speed and $\CN2$ the optical turbulence strength. Figure \ref{year_wind} shows the median vertical profile of the wind speed obtained from the ECMWF analyses during the 2003 (a) and 2004 (b) years. Dotted-lines indicate the first and third quartiles i.e. the typical dispersion at all heights. Figure \ref{year_wind} (c) shows the variability of the median profiles obtained during the two years. We can observe that from a qualitative (shape) as well as quantitative point of view (values) the results are quite similar in different years. They can therefore be considered as typical of the site. Due to the particular synoptic circulation of the atmosphere above Antarctica (the so called {\it 'polar vortex'}) the vertical distribution of the wind speed in the summer and winter time is strongly different. The wind speed has important seasonal fluctuations above $10$ km. Figure \ref{win_sum_wind} shows the median vertical profiles of the wind speed in summer (left) and winter (right) time in 2003 (top) and 2004 (bottom). We can observe that the wind speed is quite weak in the first $\sim$$10$ km from the sea-level during the whole year with a peak at around $8$ km from the sea level ($5$ km from the ground). {\bf At this height the median value is $12$ m/sec and the wind speed is rarely larger than $20$ m/sec.} Above $10$ km from the sea level, the wind speed is extremely weak during the summer time but during the winter time, it monotonically increases with the height reaching values of the order of $30$ m/sec (median) at $20$ km. 
The typical seasonal wind speed fluctuations at $5$ and $20$ km are shown in Fig.\ref{wind_free}.  
This trend is quite peculiar and different from that observed above mid-latitude sites. \newline

In order to give an idea to the reader of such differences, we show in Fig. \ref{spm_domec} the median vertical profiles of the wind speed estimated above Dome C in summer (dashed line) and winter time (full bold line) and above the San Pedro M\'artir Observatory (Mexico) in summer (dotted line) and winter time (full thin line) (Masciadri \& Egner 2004, Masciadri \& Egner 2005). San Pedro M\'artir is located in Baja California ($31.0441$ N, $115.4569$ W) and it is taken here as representative of a mid-latitude site. Above mid-latitude sites (San Pedro M\'artir - Fig.\ref{spm_domec}) we can observe that the typical peak of the wind speed at the jet-stream height (roughly 12-13 km from the sea-level) have a strong seasonal fluctuation. The wind speed is higher during the winter time (thin line) than during the summer time (dotted line) in the north hemisphere and the opposite happens in the south hemisphere. At this height, the wind speed can reach seasonal variations of the order of $30$ m/sec. Near the ground and above $17$ km the wind speed strongly decreases to low values (rarely larger than $15$ m/sec). During the winter time, the wind speed above Dome C can reach at $20$-$25$ km values comparable to the highest wind speed values obtained above mid-latitude sites at the jet-stream height (i.e. $30$ m/sec). On the other side, one can observe that, {\bf in the first $12$ km from the sea-level, the wind speed above Dome C during the winter time is weaker than the wind above mid-latitude site in whatever period of the year.} \newline
Figure \ref{seas_wind} shows, month by month, the median vertical profile of the wind speed during 2003 (green line) and during 2004 (red line). The different features of the vertical distribution of the wind speed that we have just described and attributed to the winter and summer time are more precisely distributed in the year in the following way. During December, January, February and March the median wind speed above $10$ km is not larger than $10$ m/sec. During the other months, starting from $10$ km, the median wind speed increases monotonically with different rates. September and October show the steepest wind speed growing rates. It is worth to underline the same wind speed vertical distribution appears in different years in the same month. Only during the August month it is possible to appreciate substantial differences of the median profile in the 2003 and 2004 years. This result is extremely interesting permitting us to predict, in a quite precise way, the typical features of the vertical distribution of wind speed in different months. Figure \ref{cum_wind_8_9km} shows the cumulative distribution of the wind speed at $8$-$9$ km from the sea level during each month. We can observe that, in only $20$$\%$ of cases, the wind speed reaches values of the order of $20$ m/sec during the winter time. This height ($8$-$9$ km) corresponds to the interface troposphere-tropopause above Dome C. As it will be better explained later, this is, in general, one of the place in which the optical turbulence can be more easily triggered due to the strong gradient and value of the wind speed. We remark that, similarly to what happens above mid-latitude sites, in correspondence of this interface, we find a local peak of the wind speed. In spite of this this value is much smaller above Dome C than above mid-latitude sites. We can therefore expect a less efficient production of turbulence at Dome C than above mid-latitude sites at this height. 

\subsection{Wind direction}
\label{direct}

Figure \ref{seas_wind_dir} shows, for each month, the median vertical profile of the wind direction during 2003 (green line) and during 2004 (red line).  We can observe that, during the all months, in the low part of the atmosphere the wind blows principally from the South ($\sim$$200$$^{\circ}$). In the troposphere ($1$-$11$ km) the wind changes, in a monotonic way, its direction from South to West, North/West ($\sim$$300$$^{\circ}$). In the slab characterized by the tropopause and stratosphere (above $11$ km) the wind maintains its direction to roughly $300$$^{\circ}$. Above $20$ km, during the summer time (more precisely during December, January and February) the wind changes its direction again to South. This trend is an excellent agreement
with that measured by Aristidi et al. (2005a)-Fig.6.

\subsection{Pressure}

The pressure is a quite stable parameter showing small variations during the summer and winter time above Antarctica. Figure \ref{press} shows the pressure during the summer and winter time. In this picture, we indicate the values of the pressure associated to the typical interface troposphere-tropopause above mid-latitude sites ($200$ mbar correspondent to $\sim$ $11$ km from the sea-level) and above Dome C ($300$-$320$ mbar $\sim$ $8$ km from the sea-level). As explained before, the interface between troposphere and tropopause corresponds to a favourable place in which the optical turbulence can be triggered.

\subsection{Absolute and Potential Temperature}
\label{abs_temp}

The absolute and potential temperature are fundamental elements defining the stability of the atmosphere. Figure \ref{seas_abs_temp} shows, for each month, the median vertical profile of the absolute temperature during 2003 (green line) and during 2004 (red line). Figure \ref{seas_pot_temp} shows, for each month, the median vertical profile of the potential temperature during 2003 (green line) and during 2004 (red line)\footnote{We note that, in spite of the fact that the ECMWF data are not optimized for the surface layer, the difference of the absolute temperature at the first grid point above the ground between the summer and winter time is well reconstructed by the GCMs ($\sim$ $35$$^{\circ}$ as measured by Aristidi et al.(2005))}. \newline

The value of ${\partial\ \theta} / {\partial\ z}$ indicates the level of the atmospheric thermal stability that is strictly correlated to the turbulence production. When ${\partial \ \theta}/ {\partial \ z}$ is positive, the atmosphere has high probabilities to be stratified and stable. We can observe (Fig.\ref{seas_pot_temp}) that this is observed in the ECMWF data-set and it is particularly evident during the winter time. Another way to study the stability near the ground is to analyse the ${\partial \ T} / {\partial \ z}$, i.e. the gradient of the absolute temperature. When ${\partial \ T} / {\partial \ z}$ is positive, the atmosphere is hardly affected by advection because the coldest region of the atmosphere (the heaviest ones) are already in proximity of the ground. This is a typical condition for Antarctica due to the presence of ice on the surface but it is expected to be much more evident during the winter time due to the extremely low temperature of the ice. We can observe (Fig.\ref{seas_abs_temp}) that during the winter time, ${\partial \ T}/ {\partial \ z}$ is definitely positive near the ground indicating a strongly stratified and stable conditions. All this indicates that some large wind speed gradient on a small vertical scale have to take place to trigger turbulence in the surface layer in winter time. We discuss these results with those obtained from measurements in Section \ref{rich}.\newline

A further important feature for the vertical distribution of the absolute temperature is the inversion of the vertical gradient (from negative to positive) in the free atmosphere indicating the interface troposphere-tropopause in general associated to an instable region due to the fact that ${\partial \ \theta}/ {\partial \ z}$ $\simeq$ $0$. We can observe that, above Dome C, this inversion is located at around $8$ km from the sea level during all the months. In the summer time, the median vertical profile of the absolute temperature is quite similar to the one measured by Aristidi et al. (2005a)-Fig.9. However the temperature during the winter time, above the minimum reached at $8$ km, does not increase in a monotonic way with the height but it shows a much more complex and not unambiguous trend from one month to the other with successive local minima and a final inversion from negative to positive gradients at $20$ km (May, June, July and August) and $15$ km (September and October). Considering that the regions of the atmosphere in which ${\partial \ \theta}/ {\partial \ z}$ $\simeq$ $0$ favour the instability of the atmosphere (see Section \ref{rich}), the analysis of the absolute temperature in the $8$-$25$ km range tells us that, at least from the thermal point of view, it is much more complex and difficult to define the stability of the atmosphere during the winter time than during the summer time. The Richardson number maps (Section \ref{rich}) will be able to provide us some further and more precise insights on this topic.\newline 
We finally observe that, during all the months, the vertical distribution of the absolute temperature is reproduced identically each year. 

\subsection{Richardson number}
\label{rich}

The stability/instability of the atmosphere at different heights can be estimated by the deterministic Richardson number $R_{i}$:
\begin{equation}
R_{i}=\frac{g}{\theta }\frac{\partial \theta /\partial z}{\left( \partial
V/\partial z\right) ^{2}}
\end{equation}
\noindent
where $g$ is the gravity acceleration $9.8$ m$\cdot$s$^{-2}$, $\theta$ is potential temperature and $V$ is the wind speed. The stability/instability of the atmosphere is tightly correlated to the production of the optical turbulence and it can therefore be an indicator of the turbulence characteristics above a site. The atmosphere is defined as {\it 'stable'} when $R_{i}$ $>$ $1/4$ and it is {\it 'unstable'} when $R_{i}$ $<$ $1/4$. Typical conditions of instability can be set up when, in the same region, ${\partial \ V } / {\partial \ z}$ $\gg$ $1$ and $ {\partial \ \theta}/ {\partial \ z}$ $<$ $1$ or $ {\partial \ \theta}/ {\partial \ z}$ $\sim$ $0$. Under these conditions the turbulence is triggered in strongly stratified shears. These kind of fluctuations in the atmosphere have a typical small spatial scale and can be detected by radiosoundings. When one treats meteorological parameters described at lower spatial resolution, as in our case, it is not appropriate to deal about a deterministic Richardson number. Following a statistical approach (Van Zandt et al. 1978), we can replace the deterministic $R_{i}$ with a probability density function, describing the stability and instability factors in the atmosphere provided by meteorological data at larger spatial scales. This analysis has already been done in the past by Masciadri \& Garfias (2001). Figures \ref{seas_pot_temp_grad} and \ref{seas_wind_grad} show, for each month, the gradient of the potential temperature $ {\partial \ \theta}/ {\partial \ z}$ and the square of the gradient of the wind speed $({\partial \ V } / {\partial \ z})^2$. Finally, Fig.\ref{seas_rich_inv} shows, for each month, the inverse of the Richardson number ($1/R$) over $25$ km. We show $1/R$ instead of $R$ because the first can 
be displayed with a  
better dynamic range than the second one. From a visual point of view, $1/R$ permits, therefore, to better put in evidence stability differences in different months. As explained before, with our data characterized by a low spatial resolution, we can analyze the atmospheric stability in relative terms (in space and time), i.e. to identify regions that are less or more stable then others. This is quite useful if we want to compare features of the same region of the atmosphere in different period of the year.
The probability that the turbulence is developed is larger in regions 
characterized by a large $1/R$.\newline

If, for example, we look at the $1/R$ distribution in the month of January (middle of the summer time) we can observe that, a maximum is visible at around $[2-5]$ km from the ground\footnote{We prefer to concentrate our attention to the slab [$30$ m, $\inf$) because our data-set is not optimized for study of the surface layer}, correspondent to the height at which the gradient of the wind speed has a maximum and the gradient of the potential temperature ${\partial \ \theta}/ {\partial \ z}$ $\sim$ $0$ (Fig.\ref{seas_pot_temp_grad}). The presence of both conditions is a clear indicator of instability. In the same figure, $1/R$ decreases monotonically above $5$ km indicating conditions of a general stability of the atmosphere in this region. If we compare the value of $1/R$ in different months we can easily identify two periods of the year in which the Richardson number present similar characteristics. 

During the months of December-April, $1/R$ has a similar trend over the all $25$ km. One or two peaks of $1/R$ are visible in the $[2-5]$ km region and a monotonically decreasing above $5$ km is observed. 

During the months of May-November, $1/R$ shows more complex features. At $[2-5]$ km from the ground we find a similar instability identified in the summer time but, above $5$ km, we can observe other regions of instability mainly concentrated at $12$ and $17$ km from the ground. In a few cases (in September and in October above $12$ km from the ground), the probability that the turbulence is triggered can be larger than at $[2-5]$ km. The analysis of the $R$ (or $1$/$R$) does not give us the value of the $\CN2$ at a precise height but it can give us a quite clear picture of {\it 'where'} and {\it 'when'} the turbulence has a high probability to be developed over the whole year above Dome C. \newline

Summarizing we can state that, during the whole year, we have conditions of instability in the $[2-5]$ km from the ground. We can predict the development of the turbulence but probably characterized by an inferior strength than what observed above mid-latitude sites. The wind speed at $[2-5]$ km above Dome C is, indeed, clearly weaker than the wind speed at the same height above mid-latitude sites.\newline 

In the high part of the atmosphere ($h$ $>$ $5$ km), during the summer time the atmosphere is, in general, quite stable and we should expect low level of turbulence. During the winter time the atmosphere is more instable and one should expect a higher level of turbulence than during the summer time. The optical turbulence above $10$ km would be monitored carefully in the future during the months of September and October to be sure that $\tau_{0}$ is competitive with respect to mid-latitude sites in winter time. Indeed, even a weak $\CN2$ joint to the large wind speed at these altitudes might induce important decreasing of $\tau_{0}$ with respect to the $\tau_{0}$ found above mid-latitude sites. Indeed, as can be seen in Fig.\ref{seas_wind}, the wind speed at this height can be quite strong. On the other side, we underline that this period does not coincide with the central part of the winter time (June, July and August) that is the most interesting for astronomic observations. 

We would like to stress again this concept: in this paper we are not providing absolute value of the turbulence but we are comparing levels of instabilities in different regions of the atmosphere and in different periods of the year. This status of stability/instability are estimated starting from meteorological parameters retrieved from ECMWF data-set. Considering that, as we proved once more, the meteorologic parameters are quite well described by ECMWF the relative status of stability/instability of the atmosphere represented by the Richardson number maps provided in our paper is a constrain against which measurements of the optical turbulence need to be compared. We expect that $\CN2$ measurements agree with the stability/instability properties indicated by the Richardson maps. Which is the typical seeing above the first $30$ m ? We should expect that the strength of the turbulence in the free atmosphere is larger in winter time than during the summer time. Are the measurements done so far in agreement with the Richardson maps describing the stability/instability of the atmosphere in different seasons and at different heights ?\newline

Some sites testing campaigns were organized above Dome C (Aristidi et al. 2005a, Aristidi et al. 2005b, Lawrence et al. 2004) so far employing different instruments running in different periods of the year. We need measurements provided by a vertical profiler to analyze seeing values above $30$ m. Balloons measuring vertical $\CN2$ profiles have been launched during the winter time (Agabi et al. 2006). Preliminary results indicate a seeing of $0\farcs{36}$ above the first $30$ m. Unfortunately, no measurements of the $\CN2$ vertical distribution during the summer time is available so far. Luckily, we can retrieve information on the level of activity of the turbulence in the high part of the atmosphere analysing the isoplanatic angle. 
This parameter is indeed particularly sensitive to the turbulence developed in the high part of the atmosphere. We know, at present, that the median $\theta_{0}$ measured with a GSM 
is $6\farcs{8}$ during the summer time and $2\farcs{7}$ during the winter time\footnote{We note that some discrepancies were found between $\theta_{0}$ measured by a GSM ($2\farcs{7}$) and balloons ($4\farcs{7}$) in the same period (Aristidi et al. 2005b). This should be analyzed more in detail in the future. However, in the context of our discussion, we are interested on a relative estimate i.e. on a parameter variation between summer and winter time. We consider, therefore, values measured by the same instrument (GSM) in summer and winter time.}. This means that, during the winter time, the level of the turbulence in the free atmosphere is higher than in summer time. This matches perfectly with the estimates obtained in our analysis.\newline

On the other side, a DIMM placed at $8.5$ m from the ground measured a median value of seeing $\varepsilon_{TOT}$ $=$ $0\farcs{55}$ in summer time\footnote{The seeing can reach high values in summer time as shown by Aristidi et al. 2005b} (Aristidi et al. 2005b) and $\varepsilon_{TOT}$ $=$ $1\farcs{3}$ in winter time (Agabi et al. 2006). This instrument measures the integral of the turbulence over the whole troposphere and stratosphere. The large difference of the seeing between the winter and summer time is certainly due to a general increasing of the turbulence strength near the ground in the summer-winter passage\footnote{This does not mean that one can observe some high values of $\varepsilon$ in some period of the day in summer time as shown by Aristidi et al. (2005b).}. Indeed, measurements of the seeing above $30$ m obtained with balloons and done during the winter time (Agabi et al. 2006) give a typical value of $\varepsilon _{(30m,\infty)}$ $=$ $0\farcs{36}$. Using the law:\newline
\begin{equation}
\varepsilon _{(0,30m)}=[\varepsilon _{tot}^{5/3}-\varepsilon _{(30m,\infty
)}^{5/3}]^{3/5}
\end{equation}
we can calculate that during the winter time the median seeing in the first $30$ m is equal to $\varepsilon_{(0,30m),winter}$ $=$ $1\farcs{2}$. In spite of the fact that we have no measurements of the seeing above $30$ m in summer time, we know, from the Richardson analysis shown in this paper, that the seeing in this region of the atmosphere should be weaker in summer time than in winter time. This means that the seeing above $30$ m in summer time should be smaller than $0\farcs{36}$. Knowing that the total seeing in summer time is equal to $\varepsilon_{TOT}$ $=$ $0\farcs{55}$, one can retrieve that the seeing in the first $30$ m should be smaller than $0\farcs{55}$. This means that $\varepsilon_{(0,30m),summer}$ $<$ $0\farcs{55}$ $<$ $\varepsilon_{(0,30m),winter}$ $=$ $1\farcs{2}$.\newline\newline
This means that the turbulence strength on the surface layer is larger during the winter time than during the summer time. In Section \ref{abs_temp} we said that during the winter time and near the ground, the thermal stability is larger than during the summer time. This is what the physics says and what the ECMWF data-set show but it is in contradiction with seeing measurements. The only way to explain such a strong turbulent layer near the ground during the winter time is to assume that the wind speed gradient in the first $30$ m is larger during the winter time than during the summer time. This is difficult to accept if the wind speed is weaker during the winter time than during the summer time as stated by Aristidi et al. (2005). As shown in Masciadri (2003), the weaker is the wind speed near the surface, the weaker is the gradient of the wind speed. We suggest therefore a more detailed analysis of this parameter near the surface extended over the whole year. This should be done preferably with anemometers mounted on masts or kites. This will permit to calculate also the Richardson number in the first $30$ m during the whole year and observe differences between summer and winter time. This can be certainly a useful calculation to validate the turbulence measurements. The ECMWF data-set have no the necessary reliability in the surface layer to prove or disprove these measurements.

\section{Reliability of ECMWF data}   
\label{rel}

As previously explained, measurements obtained recently above Dome C with radiosoundings (Aristidi et al. 2005a) can be useful to quantify the level of reliability of our estimates.
In Aristidi et al. (2005a) is shown (Fig.4) the median vertical profile of the wind speed measured during several nights belonging to the summer time. Figure 1 in Aristidi et al.(2005a) gives the histogram of the time distribution of measurements as a function of month. Most of measurements have been done during the December and January months. Figure \ref{mean_wind_dec_jan} (our paper) shows the vertical profile of the wind speed obtained with ECMWF data related to the December and January months in $2003$ and $2004$ (bold line) and the measurements obtained during the same months above Dome C (thin full line). E. Aristidi, member of the LUAN team, kindly selected for us only the measurements related to these two months from their sample. We note that, the ECMWF are all calculated at $00$:$00$ U.T. while the balloons were not launched at the same hour each day. Moreover, the measurements are related to 2000-2003 period while the analyses are related to the 2003-2004 period. In spite of this difference, the two mean vertical profiles show an excellent correlation. The absolute difference remains below $1$ m/sec with a mean difference of $0.7$ m/sec basically everywhere. \newline

In the high part of the atmosphere (Fig.\ref{mean_wind_dec_jan}), the discrepancy measurements/ECMWF analyses is of the order of $1.5$ $m/sec$. This is a quite small absolute discrepancy but, considering the typical wind speed value of $\sim$ $4$ $m/sec$ at this height, it gives a relative discrepancy of the order of $25$$\%$. We calculated that, assuming measurements of the seeing so far measured above Dome C and $\CN2$ profiles as shown in Section \ref{disc} (Table 2), this might induce discrepancies on the $\tau_{0}$ estimates of the order of $13$-$16$$\%$. To produce a more detailed study on the accuracy of the ECMWF analyses and measurements one should know the intrinsic error of measurements and the scale of spatial fluctuations of the wind speed at this height. No further analysis is possible for us above the Dome C to improve the homogeneity of the samples (measurements and analyses) and better quantify the correlation between them because we do not access the raw data of measurements. We decided, therefore, to compare measurements with ECMWF analyses above South Pole in summer as well as in winter time to provide to the reader further elements on the level of reliability of ECMWF analyses above a remote site such as Antarctica. Figure \ref{jan_med} (January - summer time - $12$ nights) and Fig.\ref{jj_med} (June and July  - winter time - $12$ nights) show the median vertical profiles of wind speed, wind direction and absolute temperature provided by measurements\footnote{$ftp//amrc.ssec.wisc.edu/pub/southpole/radiosonde$} and ECMWF analyses. We underline that, in order to test the reliability of ECMWF analyses, we considered all (and only) nights for which measurements are available on the whole $25$ $km$ for the three parameters: wind speed, wind direction and absolute temperature. It was observed that, during the winter time, the number of radiosounding (balloons) providing a complete set of measurements decreases. In this season it is frequent to obtain measurements only in the first $10$-$12$ $km$. Above this height the balloons blow up. To increase the statistic of the set of measurements extended over the whole $25$ $km$ we decided to take into account nights related to two months (June and July) in winter time. We can observe (Fig.\ref{jan_med}, Fig.\ref{jj_med}) that the correlation ECMWF analyses/measurements is quite good in winter as well in summer time for all the three meteorologic parameters. We expressly did not smoothed the fluctuations characterized by high frequencies of measurements. The discrepancy measurements/ECMWF analyses is smaller than $1$ $m/sec$ on the whole troposphere. It is also visible that the natural typical fluctuations at small scales of the measured wind speed is $\sim$ $1$ $m/sec$. We conclude, therefore, that a correlation measurements/ECMWF analyses within $1$ m/sec error is a quite good correlation and these data-set can provide reliable initialization data for meso-scale models.\newline
As a further output of this study we observe that, during the winter time, the wind speed above South Pole is weaker than above Dome C, particularly above $8$ km from the ground. This fact certainly affects the value of the $\tau_{0}$ placing the South Pole in a more favourable position with respect to Dome C. On the other side, we know that the turbulent surface layer is much more stronger and thicker above South Pole than Dome C. This elements also affects the $\tau_{0}$ placing Dome C in a more favourable position with respect to South Pole. Further measurements are necessary to identify which of these two elements (a larger wind speed at high altitudes above Dome C or a stronger turbulence surface layer above South Pole) more affects the $\tau_{0}$. Indeed, if typical values of $\tau_{0}$ ($1.58$ msec) in winter time (June, July and August) above South Pole are already available (Marks et al. 1999), we have not yet measurements of $\tau_{0}$ above Dome C related to the same period. Of course, if $\tau_{0}$ above Dome C will reveal to be larger than $1.58$ msec, this would mean that the stronger turbulence layer in the surface above South Pole affects $\tau_{0}$ more than the larger wind speed at high altitudes above Dome C. This study is fundamental to define the potentialities of these sites for applications to the interferometry and adaptive optics.

\section{Discussion}
\label{disc}

\begin{figure*}
\centering
\includegraphics[width=16cm]{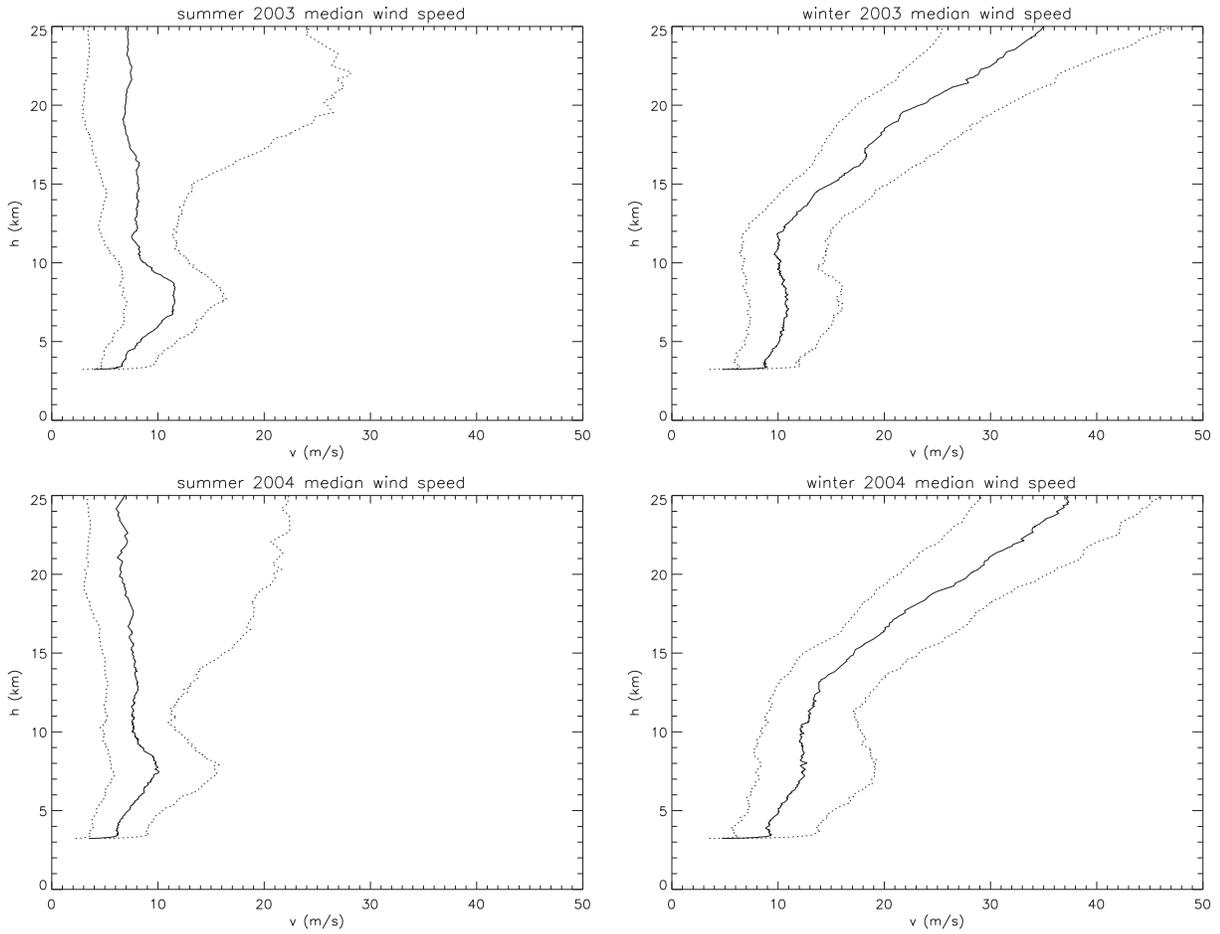}
\caption{Summer (left) and winter (right) median wind speed vertical profile estimated in 2003 (top) and 2004 (bottom). The first and third quartiles are shown with a dotted line.}
\label{win_sum_wind}
\end{figure*}

\begin{figure}
\resizebox{\hsize}{!}{\includegraphics[width=7cm,angle=-90]{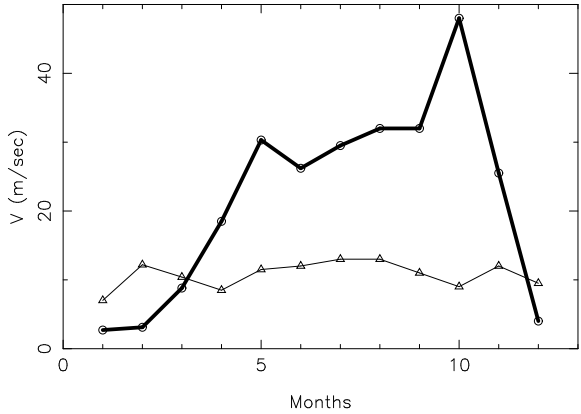}}
\caption{Seasonal trend of median wind speed estimated with ECMWF data above Dome C at $8$ km (thin line) and $20$ km (bold line). This seasonal trend shows the effect of the so called {\it 'polar vortex'}.}
\label{wind_free}
\end{figure}

\begin{figure}
\resizebox{\hsize}{!}{\includegraphics[width=7cm,angle=-90]{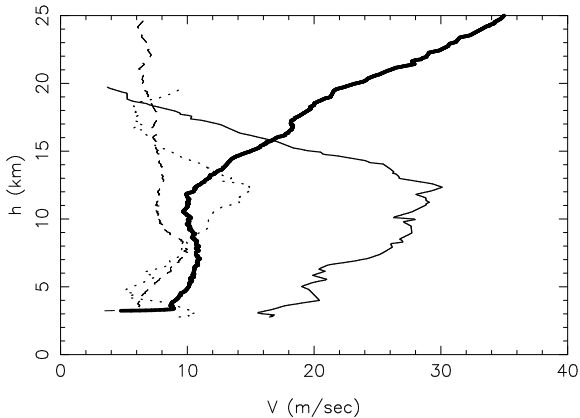}}
\caption{Comparison between the median wind speed profile estimated during the winter time (full bold line) and summer time (dashed line line) (2003) above Dome C and the median wind speed profile estimated above the San Pedro M\'artir Observatory in summer (dotted line) and winter (full thin line) time. San Pedro M\'artir is taken as representative of a mid-latitude site.}
\label{spm_domec}
\end{figure}

\begin{figure*}
\centering
\includegraphics[width=17cm]{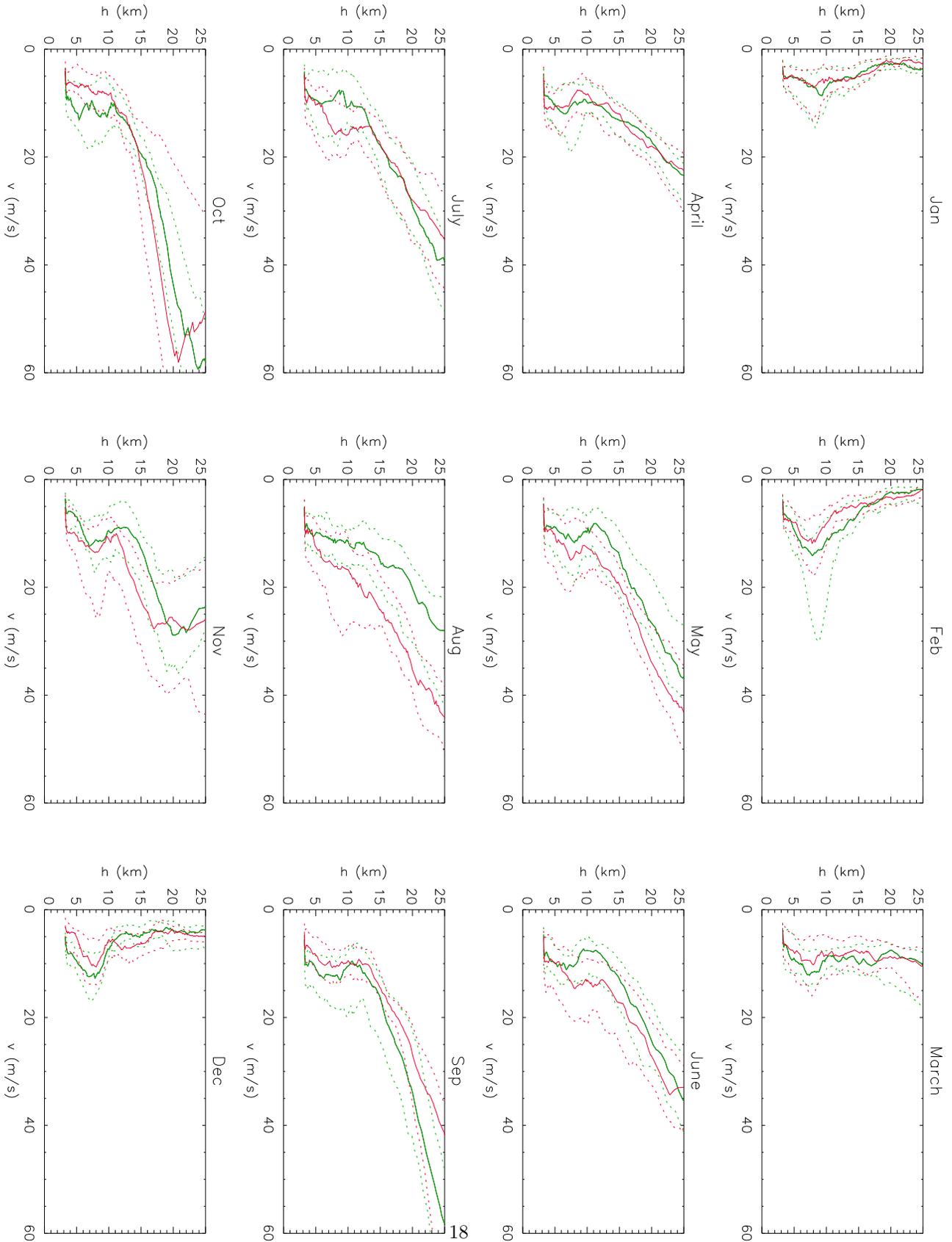}
\caption{Seasonal median {\bf wind speed} vertical profiles. Green line: year 2003. Red line: year 2004. }
\label{seas_wind}
\end{figure*}

\begin{figure*}
\centering
\includegraphics[width=17cm]{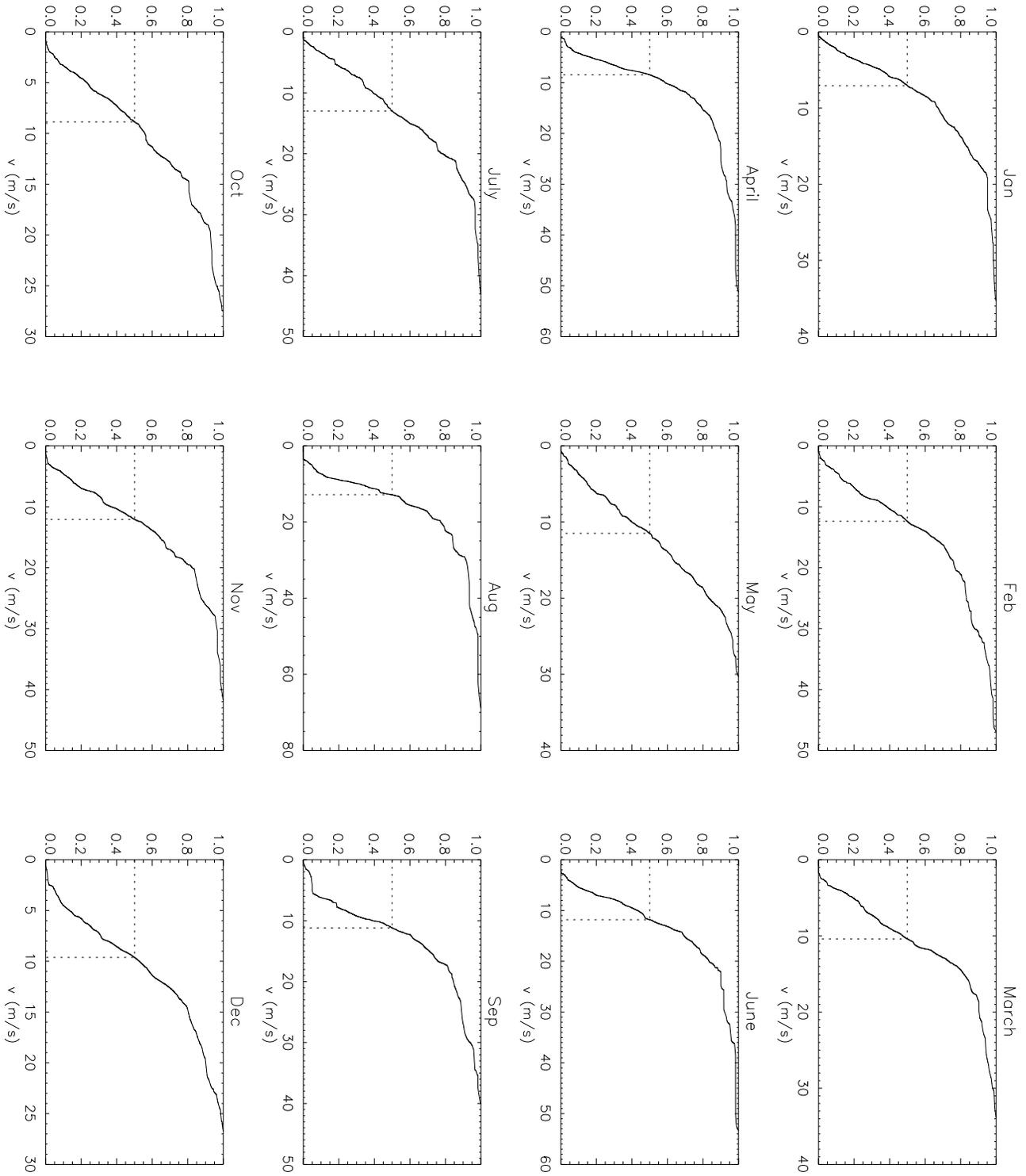}
\caption{Seasonal cumulative distribution of the wind speed in the $8$-$9$ km range from the sea level. This corresponds roughly to the tropopause height above Dome C. The pressure at this altitude is around $320$ mb.}
\label{cum_wind_8_9km}
\end{figure*}

\begin{figure*}
\centering
\includegraphics[width=17cm]{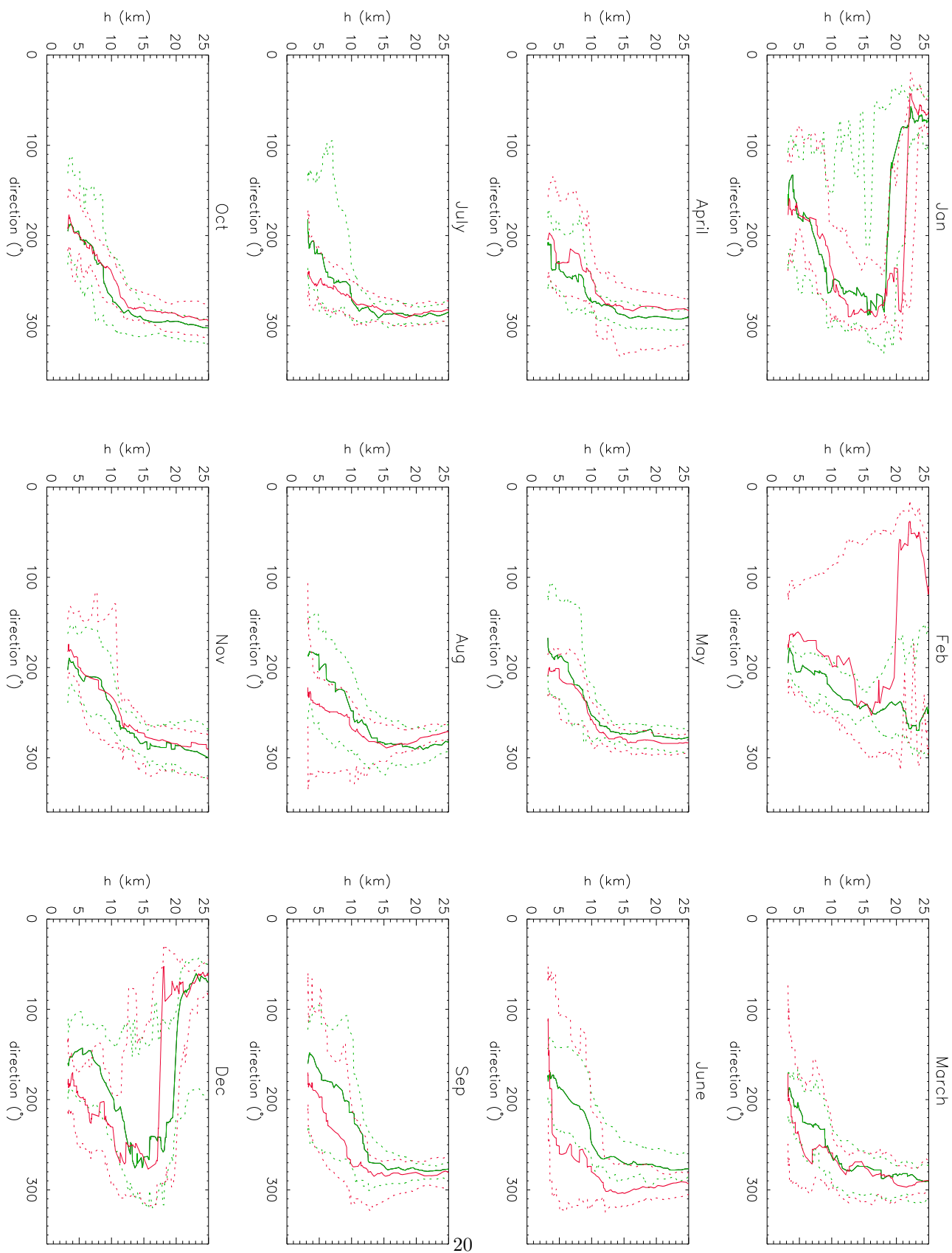}
\caption{Seasonal median {\bf wind direction} vertical profiles. Green line: year 2003. Red line: year 2004. $0$$^{\circ}$ corresponds to the North.}
\label{seas_wind_dir}
\end{figure*}

\begin{figure}
\centering
\resizebox{\hsize}{!}{\includegraphics[width=7cm]{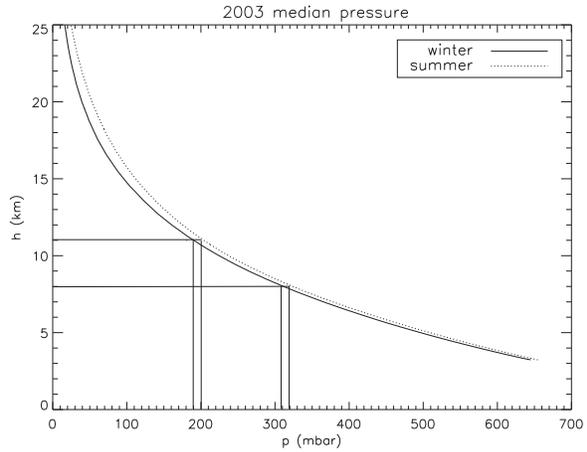}}
\caption{Atmospheric pressure in winter and summer time during the 2003. The figure shows 
the typical pressure $300$-$320$ mbar associated to the $8$ km latitude and the $190$-$200$ 
mbar associated to the $11$ km altitude.}
\label{press}
\end{figure}

\begin{figure*}
\centering
\includegraphics[width=17cm]{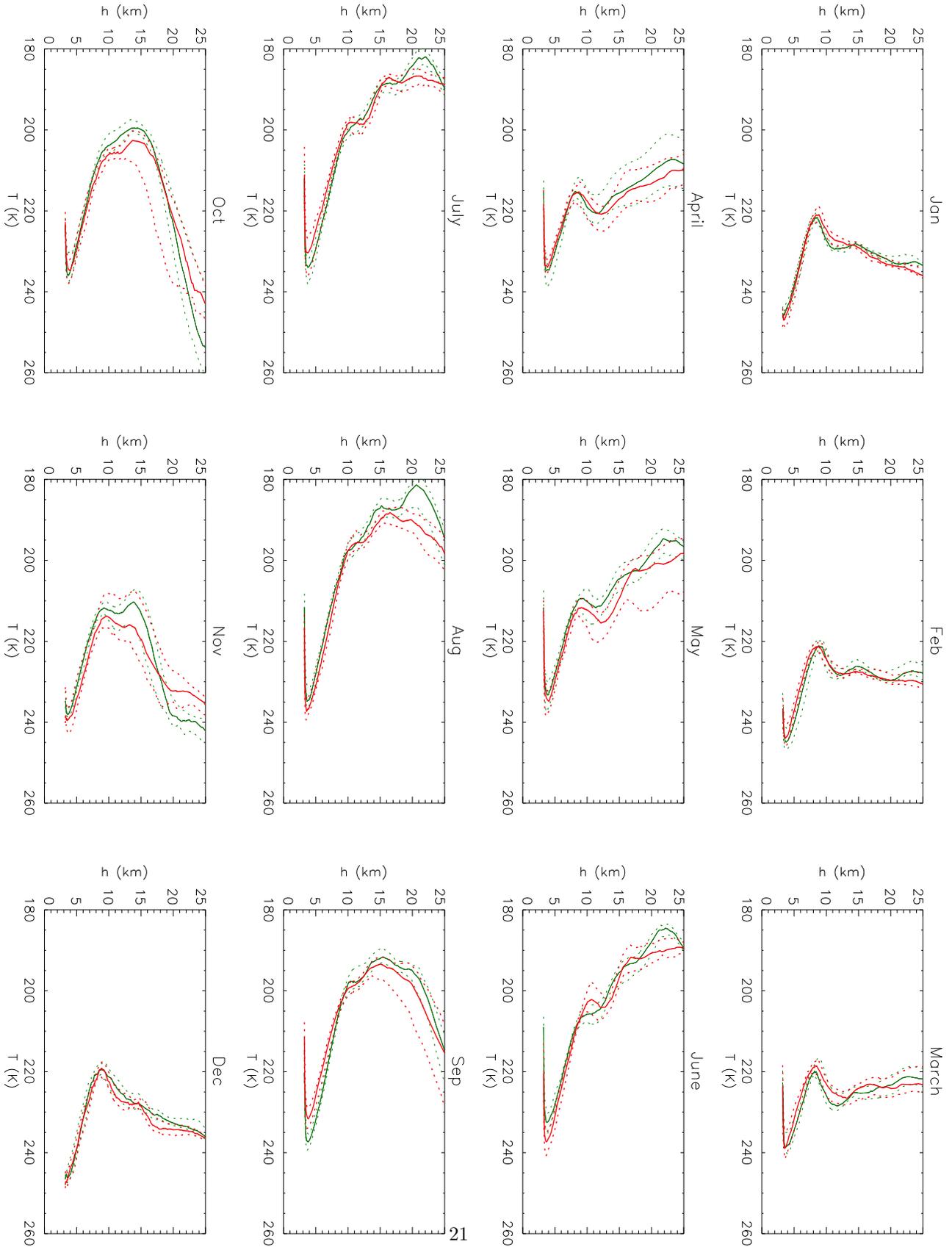}
\caption{Seasonal median {\bf absolute temperature} vertical profiles. Green line: year 2003. Red line: year 2004.}
\label{seas_abs_temp}
\end{figure*}

\begin{figure*}
\centering
\includegraphics[width=17cm]{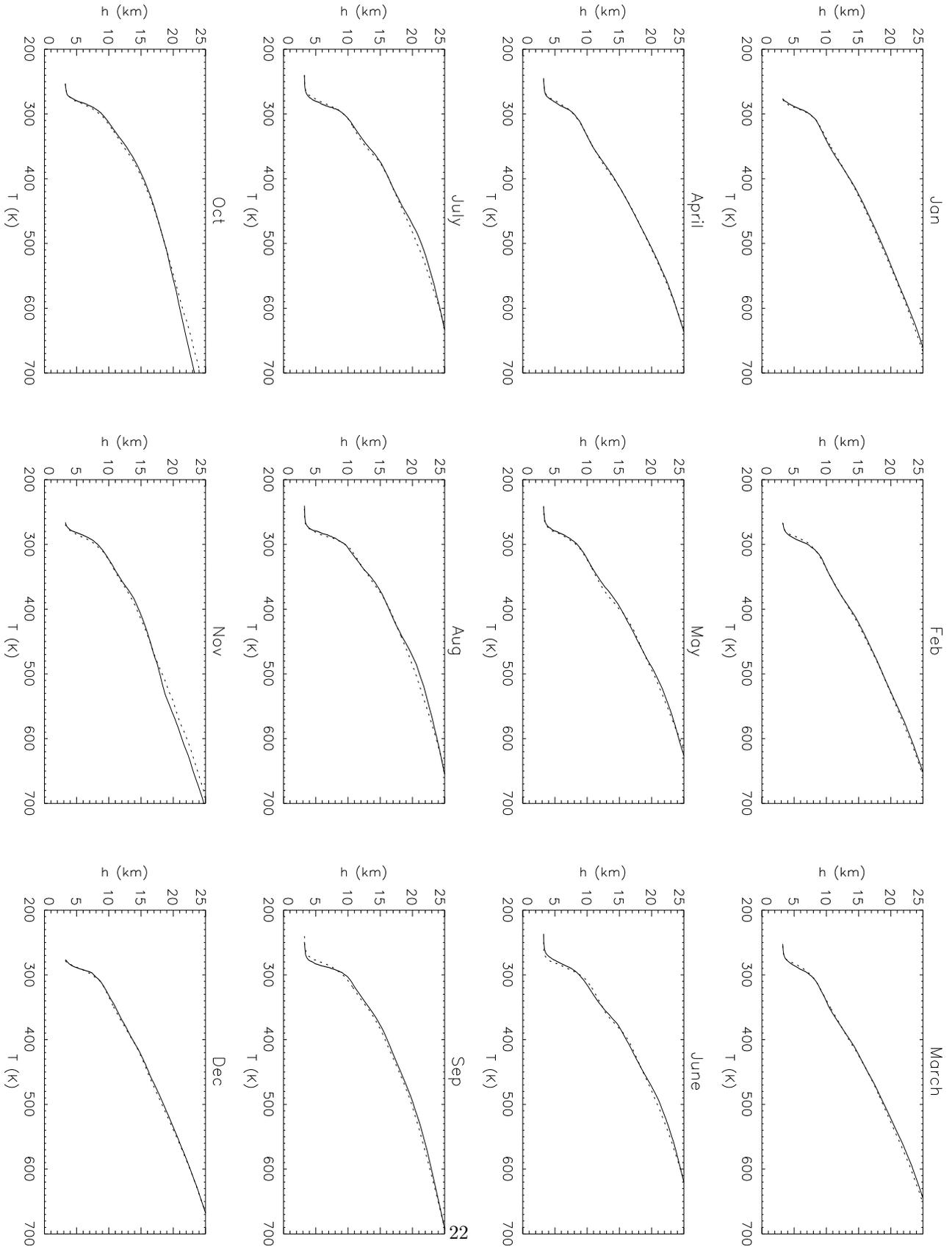}
\caption{Seasonal median {\bf potential temperature} vertical profiles. Thin full line: year 2003. Dotted line: year 2004. Dashed line: years 2003 and 2004.}
\label{seas_pot_temp}
\end{figure*}

\begin{figure*}
\centering
\includegraphics[width=17cm]{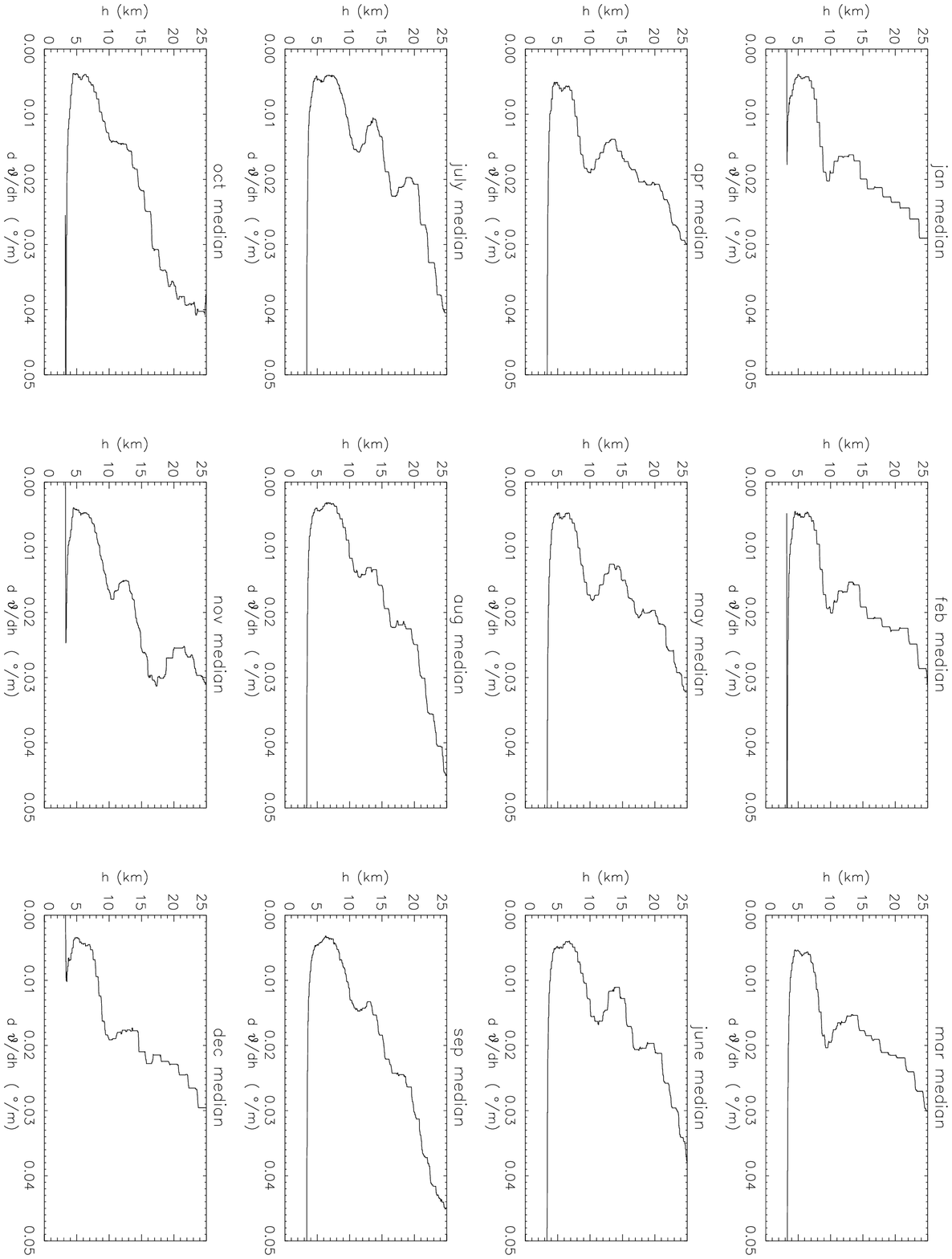}
\caption{Seasonal median {\bf $\frac{\partial \theta }{\partial h}$} vertical profile calculated with ECMWF analyses of 2003 and 2004.}
\label{seas_pot_temp_grad}
\end{figure*}

\begin{figure*}
\centering
\includegraphics[width=17cm]{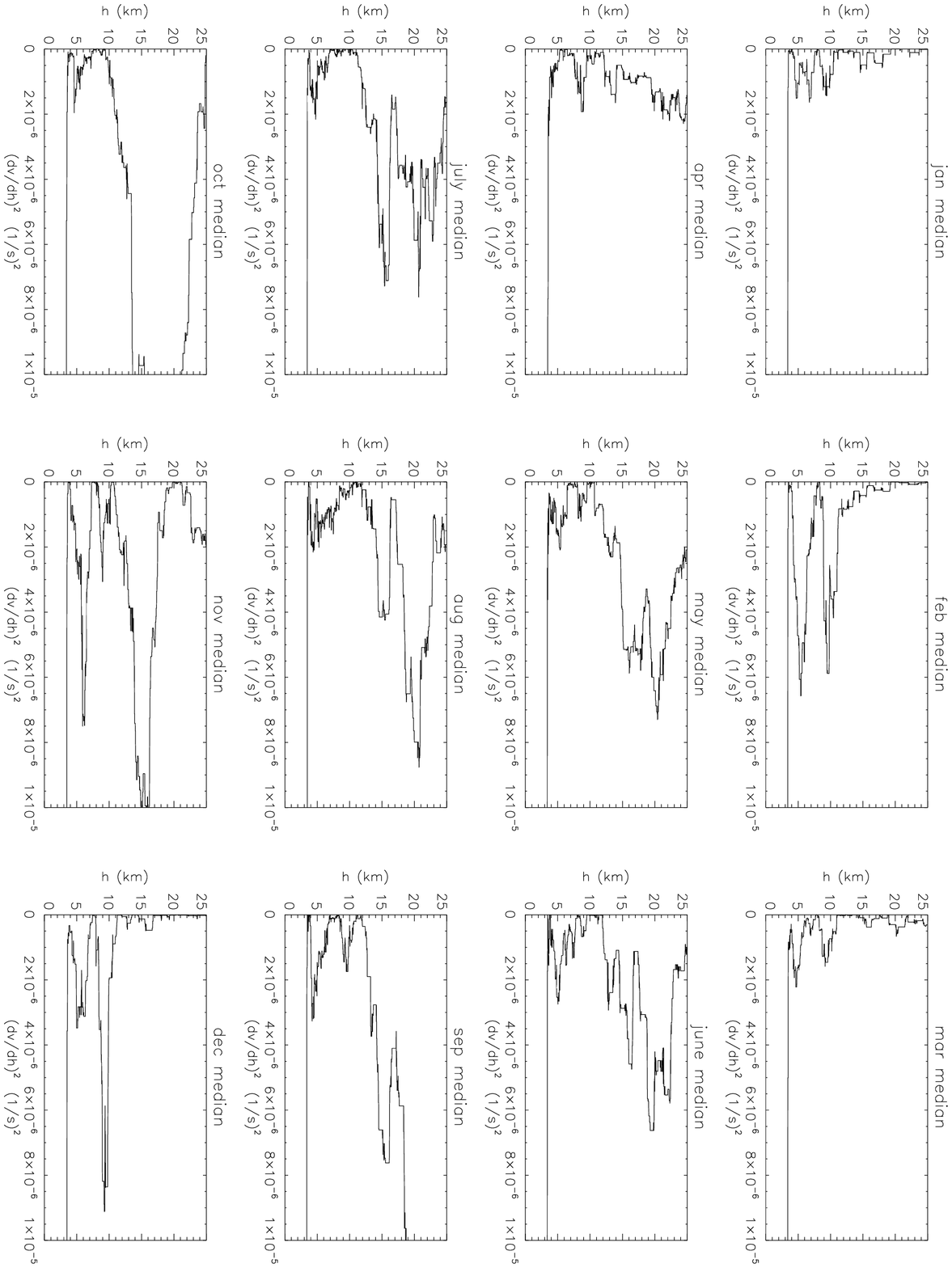}
\caption{Seasonal median {\bf ${(\frac{\partial V }{\partial h})}^2$} vertical profile calculated with ECMWF analyses of 2003 and 2004.}
\label{seas_wind_grad}
\end{figure*}

\begin{figure*}
\centering
\includegraphics[width=17cm]{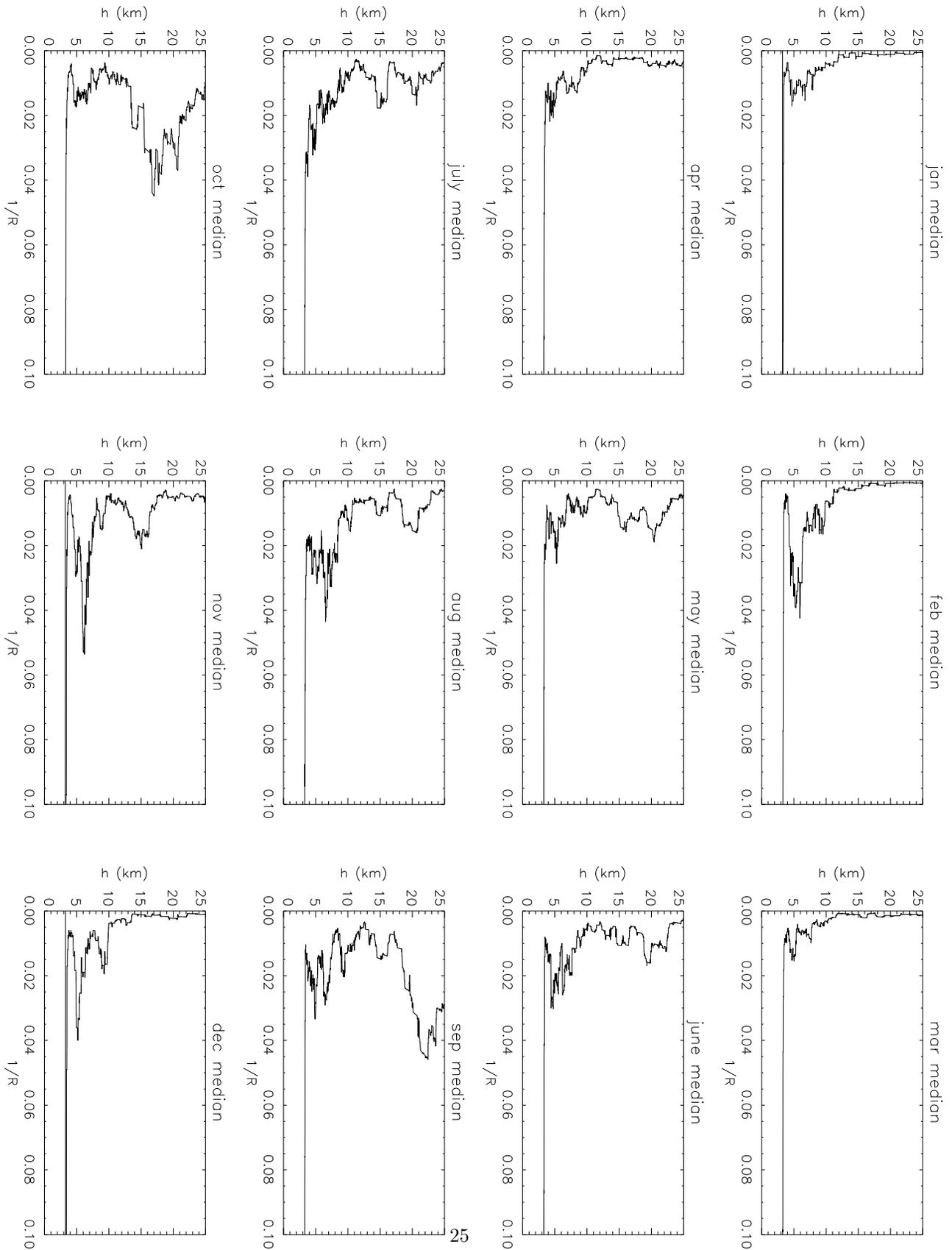}
\caption{Seasonal median {\bf 1/R - Inverse of the Richardson Number} vertical profile calculated with ECMWF analyses of 2003 and 2004.}
\label{seas_rich_inv}
\end{figure*}

\begin{figure}
\resizebox{\hsize}{!}{\includegraphics[width=6cm]{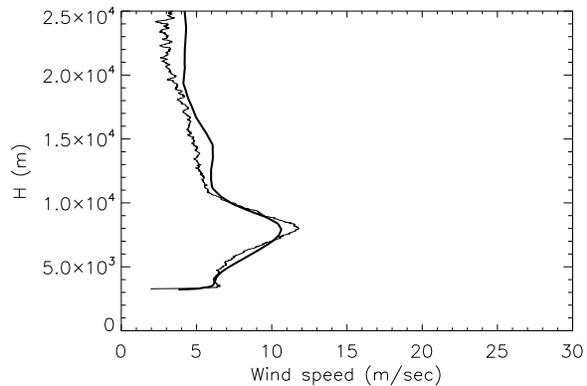}}
\caption{Mean wind speed vertical profiles measured by balloons (thin line) and provided by ECMWF data (bold line) in December and January months. Balloons measurements were published by Aristidi et al. 2005. See text for further details.}
\label{mean_wind_dec_jan}
\end{figure}

\begin{figure}
\includegraphics[width=12cm,angle=90]{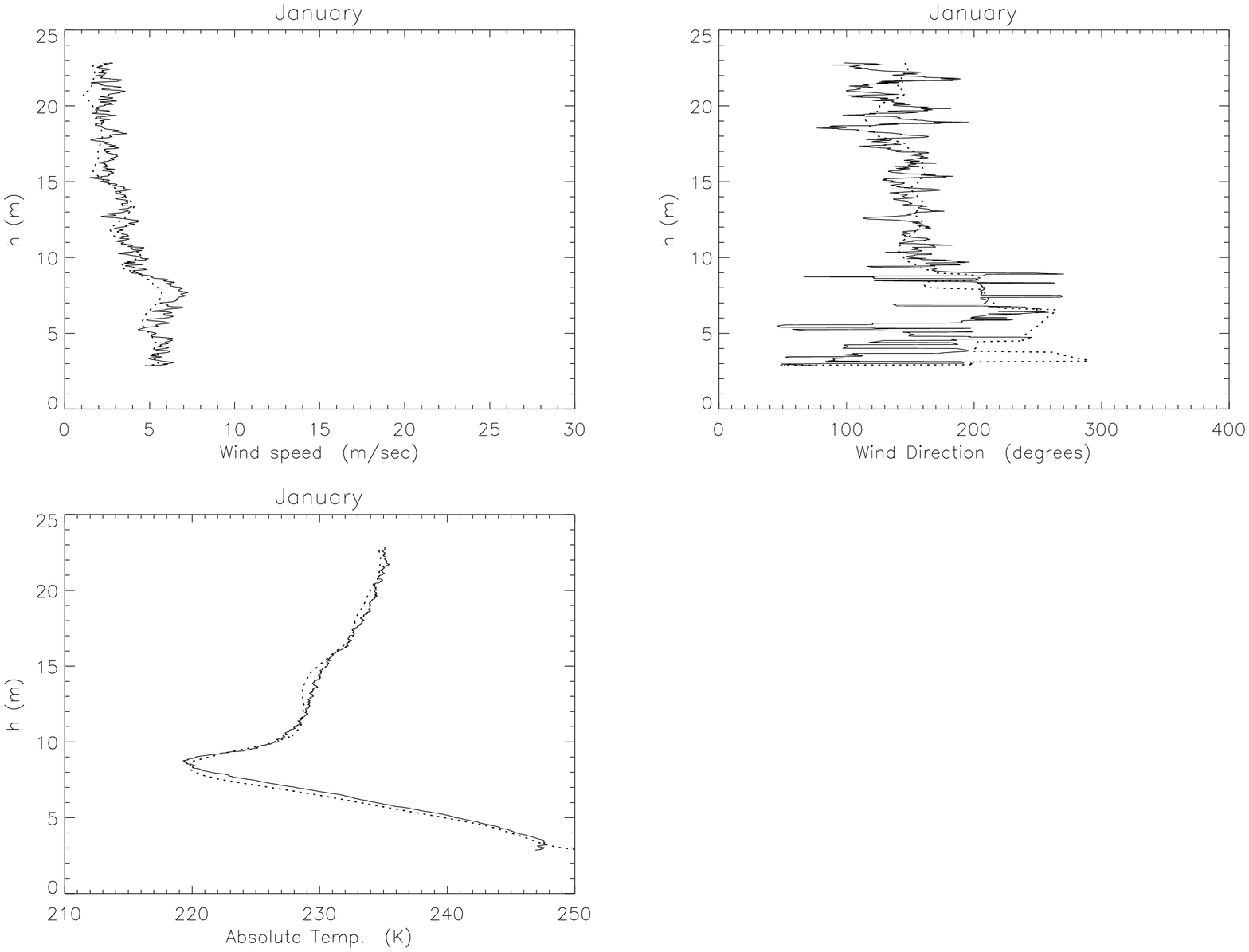}
\caption{{\bf South Pole}. ECMWF analyses (dotted line) and measurements (full line) related to $12$ nights in January 2003.}
\label{jan_med}
\end{figure}

\begin{figure}
\includegraphics[width=12cm,angle=90]{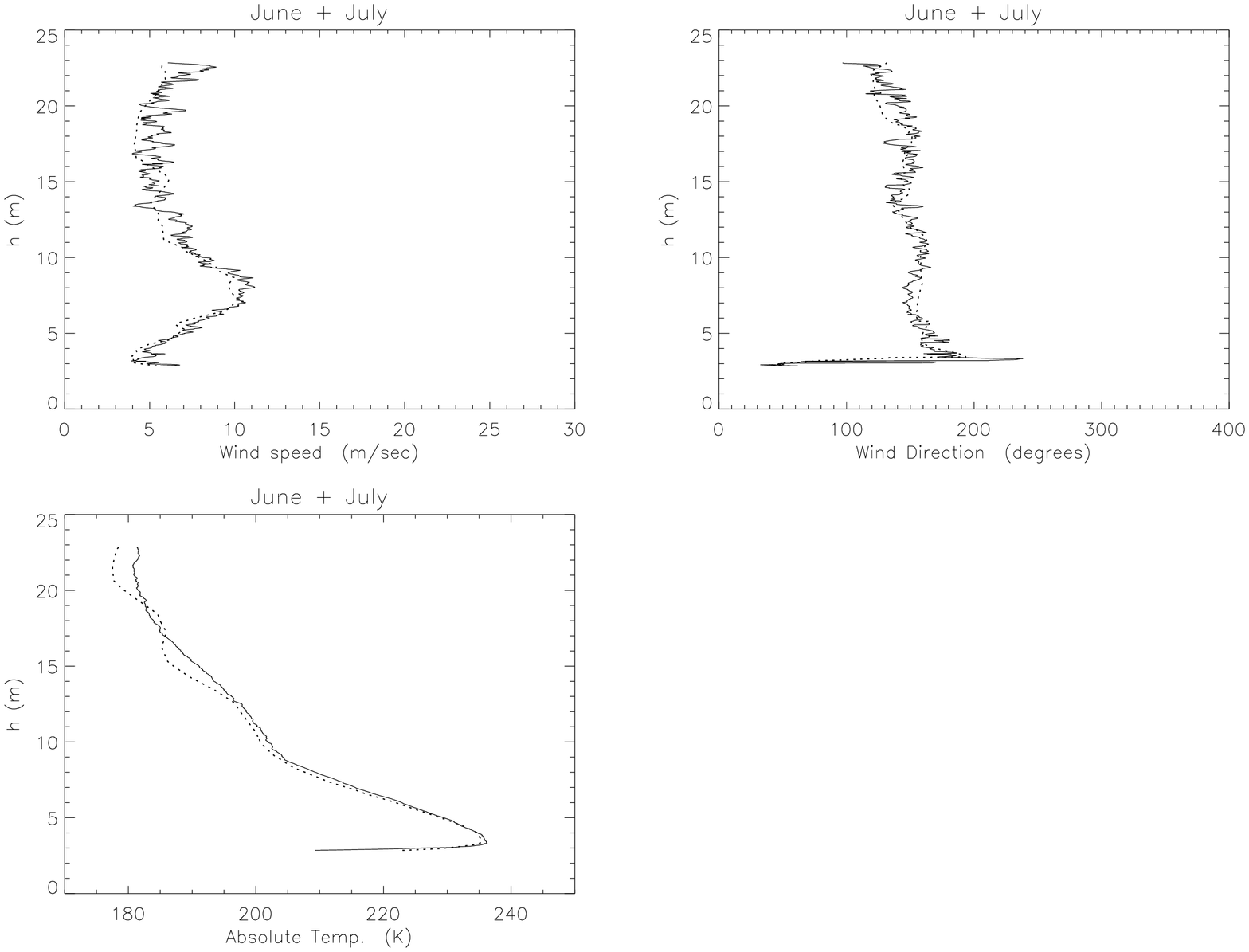}
\caption{{\bf South Pole}. ECMWF analyses (dotted line) and measurements (full line) related to $12$ nights in June and July 2003.}
\label{jj_med}
\end{figure}

We intend here to calculate the value of $\theta_{0}$, $\tau_{0}$ in the slab of atmosphere in the range [h$_{surf}$, h$_{top}$] using, as inputs, simple analytical models of the optical turbulence $\CN2$ and the median vertical profiles of the wind speed shown in Fig.\ref{win_sum_wind}. The superior limit (h$_{top}$) is defined by the maximum altitude at which balloons provide measurements before exploding and falling down. The inferior limit (h$_{surf}$) corresponds to the expected surface layer above Dome C. We define h$_{surf}$ $=$ $30$ m and h$_{ground}$ $=$ $3229$ m the Dome C ground altitude. We consider independent models with h$_{top}$ $=$ $25$ km and h$_{top}$ $=$ $20$ km. Our analysis intend to estimate typical values of some critical astroclimatic parameters ($\theta_{0}$, $\tau_{0}$) without the contribution of the first $30$ m above the iced surface. The wavefront coherence time $\tau_{0}$ is defined as Eq.(\ref{eq1}) and the isoplanatic angle $\theta_{0}$ as:
\begin{equation}
\theta _{0}=0.049\cdot \lambda ^{6/5}\left[ \int h ^{5/3}\cdot
C_{N}^{2}\left( h\right) dh\right] ^{-3/5}
\end{equation}
 

Table \ref{tab1} and Table \ref{tab2} summarize the inputs and outputs of these estimates.
\noindent
\newline
{\bf Model (A)-(F)}: The simplest (and less realistic) assumption is to consider the $\CN2$ constant over the [h$_{surf}$, h$_{top}$] range. To calculate the $\CN2$ we use three values of references: $\varepsilon$$=$$0\farcs{27}$, $\varepsilon$$=$$0\farcs{2}$ and $\varepsilon$$=$$0\farcs{1}$. We do the assumption that the $\CN2$ is uniformly distributed in the $\Delta$h where $\Delta$h$=$ h$_{top}$ - h$_{ground}$ - h$_{surf}$. We then calculate the $\CN2$ as:

\begin{equation}
C_{N}^{2}=\frac{1}{\Delta h}\left( \frac{\varepsilon }{19.96\cdot 10^{6}}\right) ^{5/3}
\label{cn2}
\end{equation}

The median vertical profiles of wind speed during the summer time in the 2003 and 2004 years (see Fig.\ref{win_sum_wind}) are used for the calculation of $\tau_{0}$.\newline\newline
{\bf Model (G)-(N)}: As discussed previously the turbulence above Dome C would preferably trigger at around $[2-5]$ km from the ground during the summer time. A more realistic but still simple model consists therefore in taking a thin layer of $\Delta$h$_{2}$$=$$100$ m thickness at $5$ km from the ground and the rest of the turbulent energy uniformly distributed in the complementary $\Delta$h$_{1}$$=$$\Delta$h - $\Delta$h$_{2}$. This model is particularly adapted to describe the $\CN2$ in summer time in which there is a well localized region of the atmosphere in which the turbulence can more easily trigger (see Section \ref{rich}). Considering the more complex morphology of the Richardson number during the winter time, we think that these simple $\CN2$ models {\bf (A)-(N)} should not well describe the turbulence vertical distribution in this season. In other worlds, we have not enough elements to assume a realistic $\CN2$ model for the winter season and we will therefore limit our analysis to the summer season. To calculate the best values of $\theta_{0}$ and $\tau_{0}$ that can be reached above Dome C we consider the realistic minimum values of C$_{N,1}^{2}$$=$10$^{-19}$ m$^{-2/3}$ (Marks et al. (1999)) given by the {\it 'atmospheric noise'}\footnote{We recall to the reader that, for values of the $\CN2$ smaller than $10$$^{-19}$m$^{(-2/3)}$ we enter in the regime of the electronic noise - see Azouit \& Vernin (2005), Masciadri \& Jabouille (2001).} and we calculate the value of the C$_{N,2}^{2}$ in the thin layer at $5$ km using Eq.(\ref{cn2}) and the following relation:

\begin{equation}
C_{N}^{2}\cdot \Delta h=C_{N,1}^{2}\cdot \Delta h_{1}+C_{N,2}^{2}\cdot
\Delta h_{2}.
\label{eq}
\end{equation}
\noindent
Aristidi et al. (2005c) measured an isoplanatic angle $\theta_{0}$$=$$6\farcs{8}$ in the summer time. Looking at Table\ref{tab1} - (Model A-F), we deduce that such a $\CN2$ uniform distribution could match with these value ($\theta_{0}$$=$$6\farcs{8}$) only in association with 
an exceptional seeing of $0\farcs{1}$. In this case, we should expect a $\tau_{0}$ of the order of $30$-$40$ msec. 
Alternatively, under the assumption of a $\CN2$ peaked at $8$ km from the sea-level (Table\ref{tab1} - (Model G-N)), a seeing of $0\farcs{2}$ would better match with the $\theta_{0}$$=$$6\farcs{8}$. In this case we should expect a $\tau_{0}$ of the order of $13$-$16$ msec. 
Summarizing we can expect the following data sets: [$\varepsilon$$=$$0\farcs{1}$, $\theta_{0}$$=$$6\farcs{8}$, $\tau_{0}$$=$$30$-$40$ msec] or [$\varepsilon$$=$$0\farcs{2}$, $\theta_{0}$$=$$6\farcs{8}$, $\tau_{0}$$=$$13$-$16$ msec]. The second one is much more realistic.\newline\newline

It is interesting to note that the $\tau_{0}$ can be quite different if one assume a seeing slightly different ($0\farcs{1}$-$0\farcs{2}$) under the hypothesis of a distribution of the $\CN2$ as described in this paper. We deduce from this analysis (joint with the discussion done in Section \ref{rich}) that the seeing above $30$ m during the summer time is probably of the order of $0\farcs{2}$ or even smaller. This means that, in the free atmosphere, the seeing should be weaker during the summer time than during the winter time (average $\varepsilon$ $=$ $0\farcs{36}$ - Agabi et al. 2006). This result well matches with our Richardson number maps. However, it would be interesting to measure the seeing in the free atmosphere during the summer time in order to better constrain the values of $\tau_{0}$. This is not evident due to the fact that radiosoundings used to measure the $\CN2$ so far can not be used to measure this parameter during the summer time. Measurements are not reliable due to fictitious temperature fluctuations experienced by the captors in this season (Aristidi, private communication). From this simple analysis we deduce reasonable values of $\theta_{0,max}$$\sim$$10-11$$\arcsec$ and a $\tau_{0,max}$$\sim$$16$ msec during the summer time under the best atmospheric conditions and the most realistic distribution of $\CN2$ in the atmosphere. We remind to the reader that some measurements of $\tau_{0}$ have already been published (Lawrence et al. 2004). Such measurements have been done just in the interface summer-winter time (April-May). Our simple $\CN2$ model is not adapted to compare estimates of $\tau_{0}$ and $\theta_{0}$ similar to those done in this Section with those measured by Lawrence et al. (2004). A more detailed information on the $\CN2$ measurements in winter time will permit in the future to verify measurements done by Lawrence et al. (2004). 

\section{Conclusion}
\label{conc}

In this paper we present a complete study of the vertical distribution of all the main meteorological parameters (wind speed and direction, pressure, absolute and potential temperature) characterizing the atmosphere above Dome C from a few meters from the ground up to $25$ km. This study employs the ECMWF {\it 'analyses'} obtained by General Circulation Models (GCM); it is extended over two years 2003 and 2004 and it provides a statistical analysis of all the meteorological parameters and the Richardson number in each month of a year. This parameter provides us useful insights on the probability that optical turbulence can be triggered in different regions of the atmosphere and in different periods of the year. The Richardson number monitors, indeed, the conditions of stability/instability of the atmosphere from a dynamic as well as thermal point of view. The main results obtained in our study are:\newline
\begin{itemize}

\item The wind speed vertical distribution shows two different trends in summer and winter time due to the {\it 'polar vortex'} circulation. In the first $8$ km above the ground the wind speed is extremely weak during the whole year. The median value at $5$ km, correspondent to the peak of the profile placed at the interface troposphere/tropopause, is $12$ m/sec. At this height the 3rd quartile of the wind speed is never larger than $20$ m/sec. Above $5$ km the wind speed remains extremely weak (the median value is smaller than $10$ m/sec) during the summer time. During the winter time the wind speed increases monotonically with the height and with an important rate reaching, at $25$ km, median values of the order of $30$ m/sec. A fluctuation of the order of $20$ m/sec is estimated at $20$ km between the summer and winter time.

\item The atmosphere above Dome C shows a quite different regime of stability/instability in summer and winter time. During the summer time the Richardson number indicates a general regime of stability in the whole atmosphere. The turbulence can be triggered preferably at [$2$-$5$] km from the ground. During the winter time the atmosphere shows a more important turbulent activity. 
In spite of the fact that the analysis of the Richardson number in different months of the year is qualitative\footnote{It does not provide a measure of the $\CN2$ profiles but the relative probability to trigger turbulence in the atmosphere.} our predictions are consistent with preliminary measurements obtained above the site in particular period of the year. Considering the good reliability of the meteorological parameters retrieved from the ECMWF analyses the Richardson maps shown here should be considered as a reference to check the consistency of further measurements of the optical turbulence in the future.   

\item With the support of a simple model for the $\CN2$ distribution, the Richardson number maps and the wind speed vertical profile we calculated a best $\theta_{0,max}$$\sim$$10$\arcsec$-11$$\arcsec$ and $\tau_{0,max}$$\sim$$16$ msec above Dome C during the summer time.

\item The vertical distribution of all the meteorological parameters show a good agreement with measurements. This result is quite promising for the employing of the ECMWF analyses as initialization data for meso-scale models. Besides, it opens perspectives to
employ ECMWF data for a characterization of meteorologic parameters extended over long timescale.

\end{itemize}

\acknowledgments

Data-sets from MARS catalog (ECMWF) were used in this paper. 
This study was supported by the 
Special Project (spdesee) - ECMWF-\small{http://www.ecmwf.int/about/special}\_\small{projects/index.html}. 
We thanks the team of LUAN (Nice - France): Jean Vernin, Max Azouit, Eric Aristidi, Karim Agabi and Eric Fossat for
kindly providing us the wind speed vertical profile published in Aristidi et al. (2005a).
We thanks Andrea Pellegrini (PNRA - Italy) for his kindly support to this study.
This work was supported, in part, by the Community's Sixth Framework Programme and the Marie Curie Excellence Grant (FOROT).

\clearpage


%

\clearpage

\clearpage

\begin{table*}
\begin{center}
\begin{tabular}{ccccccc}
\hline
\hline
\noalign{\smallskip}
  &  &  &  &  & sum-2003 & sum-2004  \\
Models & h$_{top}$ & $\see$ & $\CN2$  & $\theta_{0}$ & $\tau_{0}$ & $\tau_{0}$  \\
    & (km) & (arcsec) & m$^{(-2/3)}$ & (arcsec) & (msec) &  (msec) \\
\hline
\noalign{\smallskip}
Model A  & 25 & 0.27 & 3.53$\cdot$10$^{-18}$ &  1.95& 14.00 &  15.38    \\
Model B  & 25 & 0.2 & 2.14$\cdot$10$^{-18}$ &  2.63& 18.91 &  20.77 \\
Model C  & 25 & 0.1 & 6.74$\cdot$10$^{-19}$ &  5.26& 37.83 &  41.54   \\
\hline
Model D  & 20 & 0.27 & 4.58$\cdot$10$^{-18}$ & 2.53& 13.52&  14.87   \\
Model E  & 20 & 0.2 & 2.78$\cdot$10$^{-18}$ &  3.41& 18.25&  9.89   \\
Model F  & 20 & 0.1 & 8.76$\cdot$10$^{-19}$ &  6.82& 36.49&  40.11   \\
\noalign{\smallskip}
\hline
\hline
\noalign{\smallskip}
\end{tabular}
\caption{Isoplanatic angle $\theta_{0}$, wavefront coherence time $\tau_{0}$ during the summer time in 2003 and 2004.}
\label{tab1}
\end{center}
\end{table*}
\begin{table*}
\begin{center}
\begin{tabular}{ccccccc}
\hline
\hline
\noalign{\smallskip}
  &  &  &  &  & sum-2003 &  sum-2004  \\
Models & h$_{top}$ & $\see$ & $\CNb$  & $\theta_{0}$ & $\tau_{0}$ & $\tau_{0}$  \\
    & (km) & (arcsec) & m$^{(-2/3)}$ & (arcsec) & (msec) & (msec) \\
\hline
\noalign{\smallskip}
Model G  & 25 & 0.27 & 7.46$\cdot$10$^{-16}$ & 4.60& 10.17 & 9.07  \\
Model H  & 25 & 0.2 & 4.40$\cdot$10$^{-16}$ & 6.03& 13.87 &  12.14 \\
Model I  & 25 & 0.1 & 1.25$\cdot$10$^{-16}$ & 10.18& 28.34 &  33.00   \\
\hline
Model L  & 20 & 0.27 & 7.51$\cdot$10$^{-16}$ & 4.73 & 10.15&  11.89\\
Model M  & 20 & 0.2 & 4.49$\cdot$10$^{-16}$ & 6.32 & 13.75 & 16.08   \\
Model N  & 20 & 0.1 & 1.30$\cdot$10$^{-16}$ & 11.65 & 28.02&  32.67 \\
\noalign{\smallskip}
\hline
\hline
\noalign{\smallskip}
\end{tabular}
\caption{Isoplanatic angle $\theta_{0}$, wavefront coherence time $\tau_{0}$ during the summer time in 2003 and 2004. 
$\CNd$$=$10$^{-19}$m$^{(-2/3)}$ in all the models.}
\label{tab2}
\end{center}
\end{table*}

\end{document}